\documentclass[preprint,aps]{revtex4}

\usepackage{graphicx}
\usepackage{dcolumn}
\usepackage{bm}
\usepackage{epsf}
\usepackage{epsfig}

\usepackage{amsmath}
\usepackage{amssymb}
\usepackage{pifont}
\usepackage{srcltx}

\newcommand{\be}{\begin{eqnarray}}

\newcommand{\Eq}[1]{Eq.~(\ref{#1})}
\newcommand{\Eqs}[2]{Eqs.(\ref{#1},\ref{#2})}
\newcommand{\Eqss}[3]{Eqs.(\ref{#1},\ref{#2},\ref{#3})}
\newcommand{\Eqsss}[2]{Eqs.(\ref{#1}--\ref{#2})}
\newcommand{\ur}[1]{(\ref{#1})}
\newcommand{\urs}[2]{(\ref{#1},\ref{#2})}
\newcommand{\urss}[3]{(\ref{#1},\ref{#2},\ref{#3})}

\newcommand{\beq}{\begin{equation}}
\newcommand{\eeq}{\end{equation}}

\newcommand{\la}[1]{\label{#1}}
\newcommand{\bea}{\begin{eqnarray}}
\newcommand{\eea}{\end{eqnarray}}
\newcommand{\beqa}{\begin{eqnarray}}
\newcommand{\eeqa}{\end{eqnarray}}
\newcommand{\ba}{\begin{array}}
\newcommand{\ea}{\end{array}}

\newcommand{\n}{\nonumber}
\newcommand{\nn}{\nonumber}
\newcommand{\Tr}{{\rm Tr}}

\newcommand{\eps}{\epsilon^{\kappa\lambda\mu\nu}}
\newcommand{\bydef}{\stackrel{d}{=}}

\renewcommand{\(}{\left(}
\renewcommand{\)}{\right)}
\renewcommand{\[}{\left[}
\renewcommand{\]}{\right]}

\def\appendix{\par
\setcounter{section}{0}
\setcounter{subsection}{0}
\setcounter{equation}{0}

\def\thesection{Appendix}
\def\theequation{\Alph{section}.\arabic{equation}}}

\begin{document}

\title{\bf Phase transitions in spinor quantum gravity on a lattice}

\author{Alexey A. Vladimirov$^1$ and Dmitri Diakonov$^{2,3}$}
\vskip 1true cm

\affiliation{$^1$ Ruhr-Universit\"at Bochum, Bochum D-44780, Germany\\
$^2$ Petersburg Nuclear Physics Institute, Gatchina 188300, St. Petersburg, Russia \\
$^3$ St. Petersburg Academic University, St. Petersburg 194021, Russia}



\begin{abstract}

We construct a well-defined lattice-regularized quantum theory formulated in terms of fundamental
fermion and gauge fields, the same type of degrees of freedom as in the Standard Model. The theory
is explicitly invariant under local Lorentz transformations and, in the continuum limit, under
diffeomorphisms. It is suitable for describing large nonperturbative and fast-varying fluctuations
of metrics. Although the quantum curved space turns out to be on the average flat and smooth
owing to the non-compressibility of the fundamental fermions, the low-energy Einstein limit
is not automatic: one needs to ensure that composite metrics fluctuations propagate to long distances
as compared to the lattice spacing. One way to guarantee this is to stay at a phase transition.

We develop a lattice mean field method and find that the theory typically has several phases
in the space of the dimensionless coupling constants, separated by the 2$^{\rm nd}$
order phase transition surface. For example, there is a phase with a spontaneous breaking of chiral
symmetry. The effective low-energy Lagrangian for the ensuing Goldstone field is explicitly
diffeomorphism-invariant. We expect that the Einstein gravitation is achieved at the phase
transition. A bonus is that the cosmological constant is probably automatically zero.\\




\end{abstract}

Dated: September 21, 2012

\maketitle

\section{Introduction}

We live in a world with fermions, and they must be included into General Relativity.
The standard way one couples Dirac fermions to gravity is via the Fock--Weyl
action~\cite{Fock:1929vt,Weyl:1929fm}: Fermions interact with the frame field $e^A_\mu$
(also known as {\it vierbein, rep\`ere} or {\it tetrad}) and with the spin connection $\omega_\mu^{AB}$
being the gauge field of the local Lorentz group. The frame and the spin connection are
{\it a priori} independent field variables. The bosonic part of the action has to be written
through $e_\mu$ and $\omega_\mu$ accordingly. This is known for the last 90 years as Cartan's
formulation of General Relativity~\cite{Cartan:1923}. Speaking generally, it is distinct from
the classic Einstein--Hilbert formulation based on the Riemann geometry, since it allows for
a nonzero torsion. We stress that the presence of fermions in Nature forces us to make
a definite choice in favor of the Cartan, as contrasted to the Riemann geometry.

In practice, however, it is hardly possible to detect the difference. In the leading order in the
gradient expansion of the gravitational action written down in terms of $e_\mu$ and $\omega_\mu$,
the saddle-point equation for $\omega_\mu$ says that torsion is on the average zero. Therefore,
Cartan's theory reduces to that of Einstein.

In the next order in $p^2/M_{\rm P}^2$ where $M_{\rm P}$ is the Planck mass and $p$ is the
characteristic momentum, a four-fermion contact interaction appears from integrating out torsion.
Its strength is many orders of magnitude less than that of weak interactions~\cite{Freidel:2005sn}
therefore this correction will hardly be detected any time soon in the laboratory. In principle,
it modifies {\it e.g.} the Friedman cosmological evolution equation that follows from the purely
Riemannian approach. However the correction remains tiny as long as fermions in the Universe have
Fermi momentum or temperature that are much less than $M_{\rm P}$~\cite{Diakonov:2011fs}. If they
reach that scale such that the four-fermion correction becomes of the order of the leading
stress-energy term, the theory itself fails since the gradient expansion~\cite{Diakonov:2011fs,Baekler:2011jt}
from where it has been derived, becomes inapplicable. There is no agreed
upon idea how the theory looks like at the Planck scale; in particular, quantum gravity effects
are supposed to set up there.

Being indistinguishable from Einstein's equation in the range where observations are performed,
Cartan's theory, however, has a critical feature when one attempts to quantize it. The bosonic
part of the action is written in terms of $e_\mu$ and $\omega_\mu$. To preserve the required
general covariance or invariance under the change of coordinate system, called diffeomorphism,
any action term is necessarily odd in the antisymmetric Levi-Civita tensor $\eps$. That makes
all possible diffeomorphism-invariant action terms not sign-definite~\cite{Diakonov:2011im}.

The simplest example is the invariant volume itself or the cosmological term, $\int d^4x\,\det(e)$.
If the frame field is allowed to fluctuate, as supposed in quantum gravity, the sign of $\det(e)$
can continuously change from positive to negative or {\it vice versa}. Of course, $\det(e)=0$
is a singularity where the curved space effectively looses one dimension but it is not possible
to forbid such local happenings in the world with a fluctuating metric, see the illustration in Fig.~1,
left. Moreover, if $\det(e)$ goes to zero linearly in some parameter $t$, it {\em has} to change sign
by continuity, see Fig.~1, right. The same is true for any diffeomorphism-invariant action term.
\vskip 0.5true cm

\begin{figure}[htb]
\includegraphics[width=0.45\textwidth]{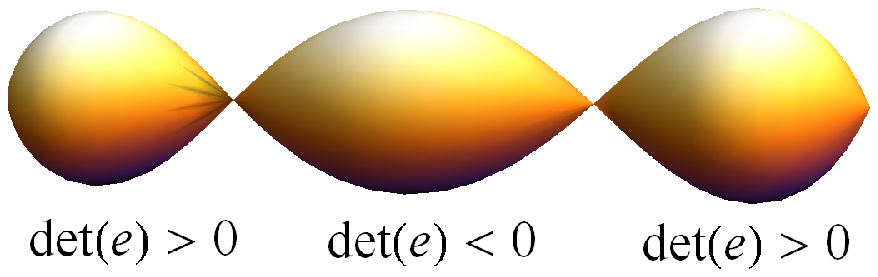}\hspace{1cm}
\includegraphics[width=0.4\textwidth]{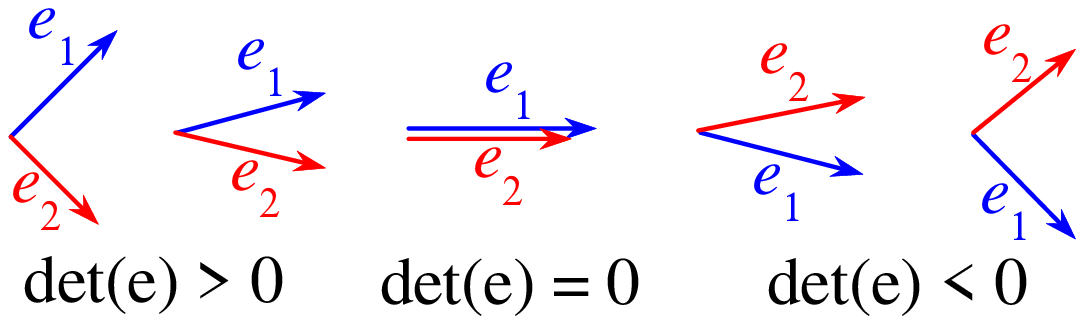}

\caption{{\it Left}:  An example of a space with alternating sign of $\det(e)$;
{\it Right}: $\det(e)$ changes sign by continuity of the frame field.}
\end{figure}


In the standard Riemannian formulation, one writes the invariant integration measure
with the help of $\sqrt{\det(g)}$ where $g_{\mu\nu}=e^A_\mu e^A_\nu$ is the metric tensor, hence
$\det(g)=(\det(e))^2$ is sign-definite. Its square root, however, should be understood as
$\sqrt{\det(g)}=\det(e)$ and can have any sign. If it passes through zero it changes sign by
continuity~\cite{Fyodorov:2008}.

This fundamental pathology of any diffeomorphism-invariant quantum theory has not been stressed before,
probably for two reasons. First, one commonly deals with the perturbative quantization about flat or
{\it e.g.} de Sitter metric such that the main concern is the absence of runaway fluctuations from that
point only. However, when quantizing gravity, one has to be concerned with large non-perturbative
fluctuations as well. Second, usually Minkowski space-times are considered where the integration measure
$\exp(i\,{\rm Action})$ is oscillating anyway independently of the action sign. However, a theory
with a sign-indefinite action in Euclidian space where the weight is $\exp(-{\rm Action})$ is usually
fundamentally sick also in Minkowski space. An illustration is provided by the scalar $\phi^3$ theory, see Fig.~2.
Perturbation theory exists there in the usual sense near $\phi=0$. However, if in Euclidian space the
theory does not exist, in Minkowski space one cannot define properly the non-perturbative Feynman propagator.
There will be also other pathologies related to the possibility of tunneling to a bottomless state.

\begin{figure}[htb]
\begin{minipage}[]{.99\textwidth}
\includegraphics[width=0.3\textwidth]{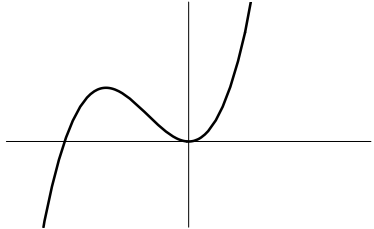}
\end{minipage}
\caption{The $\phi^3$ theory is fundamentally sick both in Euclidean space where
it is unbounded, and in Minkowski space where it can tunnel to a bottomless state.}
\label{fig:1}
\end{figure}

In gravity theory, Euclidian formulation has its own right, for example in problems related to
thermodynamics and to tunneling, like in the Hawking radiation problem where paradoxes are encountered just
because we do not know how to quantize Euclidian gravity. If a theory is well defined for Euclidian signature,
it is usually possible to Wick-rotate it to the Minkowski world. Therefore, for clearness we shall discuss
here Euclidian gravity.

Any diffeomorphism-invariant action, with any number of derivatives, is not sign-definite in Euclidian space
and hence cannot serve to define quantum gravity non-perturbatively.

At this time, we see only one way to overcome the sign problem, and that is to use in part fermionic
variables in formulating quantum gravity microscopically, rather than only bosonic ones. Integrals over
anticommuting Grassmann variables are well defined irrespectively of the overall sign in the exponent of
a fermionic action. The reason is that in fermionic integrals introduced by Berezin~\cite{Berezin:1966}
one actually picks up only certain finite order in the Taylor expansion of the exponent of the action,
such that the overall sign does not matter. One calls it {\em spinor quantum gravity}. It has been advocated
by Akama~\cite{Akama:1978pg}, Volovik~\cite{Volovik:1990} and recently in a series of papers by
Wetterich~\cite{Wetterich:2005yi,Wetterich:2011yf,Wetterich:2011xv,Wetterich:2012qu} on other grounds.

More specifically, we suggest~\cite{Diakonov:2011im} (see also Ref.~\cite{Obukhov:2012je})
that at the fundamental, microscopic level gravity theory is defined as a theory of certain
fundamental anticommuting spinor fields $\psi^\dagger,\psi$. We wish to preserve
local gauge Lorentz symmetry exactly at all stages, and for that we need the explicit connection field
$\omega_\mu$. The frame field $e_\mu$ and the metric tensor $g_{\mu\nu}$ will be composite fields
making sense only at low energies. The basic independent variables will be $\psi^\dagger,\psi$
and the gauge field $\omega_\mu$, the same type of degrees of freedom as in the Standard Model.
We believe that using the same type of variables as in the Standard Model will help to unify all
interactions~\cite{Diakonov:2011im}. As far as only gravity is concerned, the fundamental spinor fields
$\psi^\dagger,\psi$ may or may not be related to the fundamental matter fields. We introduce the main
building blocks of the theory in Section II.

A quantum field theory is well defined if it is regularized in the ultraviolet.
We shall regularize spinor quantum gravity by introducing simplicial lattice (made of triangles in 2 dimensions,
tetrahedra in 3 dimensions, 5-cells or pentachorons in 4 dimensions, {\it etc.}) covering an abstract space,
such that the simplex vertices are characterized and counted by integers $i$. Only the topology of this abstract
number space matters, {\it e.g.} the number of nearest neighbors, {\it etc.}

Each vertex $i$ in the number space corresponds to the real world coordinate by a certain map $x^\mu(i)$.
Diffeomorphism-invariance means that the theory should not depend on the coordinates $x^\mu(i)$ we ascribe
to the vertices. Also, in the continuum limit (implying slowly varying fields) the action should be of the form
$\int d^dx\,{\cal L}(x)$ and invariant under diffeomorphisms, $x^\mu \to x'^\mu(x)$. The integration
measure over the fields in the path integral formulation should be also diffeomorphism-invariant.
In addition, we require exact gauge invariance under local Lorentz transformations. We build fermionic
actions satisfying these conditions in Section III, and regularize them by putting on a lattice in Section IV.

After constructing a completely well-defined lattice-regularized quantum theory, the next question to address is
whether the continuum limit can be achieved and whether it reduces to the Einstein--Cartan theory in the
low-energy limit. The continuum limit is obtained when and if field correlations spread over a large distance
in lattice units. The trouble is that the quantum theory one deals with in this approach is a typical
strong-coupling theory where most of the correlations die out over a few lattice cells. Contrary to the standard
lattice gauge theory where long-range correlations are ensured by simply taking the weak-coupling limit
$\beta\to\infty$, in spinor quantum gravity there is no obvious handle to make the correlations long-ranged.

The main trick and the invention of this paper is to ensure long-range correlations by adjusting the
bare dimensionless coupling constants to the point (or line, or surface) where the theory undergoes
a phase transition of the 2$^{\rm nd}$ kind. At such point, all correlation functions become long-ranged,
and the Einstein theory will be guaranteed in the low-momenta limit by the inherent diffeomorphism invariance.

Second-order phase transitions occur in theories where there is an order parameter, usually related to
the spontaneous breaking of a continuous symmetry. Our primary goal in this project is to demonstrate
that second-order phase transitions are typical in the kind of diffeomorphism-invariant theories we consider.
We develop in Section V an original mean field method well suited for the search of the phase transitions, and check
its accuracy in the Appendix where it is probed in an exactly solvable model, with very satisfactory results.
Using this method we unveil the phase diagram of a generic 2-dimensional lattice spinor gravity, in the
space of the bare coupling constants, Section VI. The model has two continuous symmetries: the $U(1)$ chiral symmetry
and the $U(1)$ symmetry related to the fermion number conservation; both can be in principle spontaneously broken.

It turns out that there is a range of bare couplings where the fermionic lattice system experiences
spontaneous breaking of chiral symmetry. In the particular model we studied we did not observe spontaneous
breaking of the fermion number, however it can happen in other models. This is an interesting finding
{\it per se} but it may be also of use in the attempts to unify quantum gravity with the Standard Model.
On this route, one expects one or several spontaneous breakups of continuous symmetries.

The 2-dimensional model we consider in some detail has certain nice features. First, the physical (invariant) volume
$\langle V\rangle$ is extensive {\it i.e.} proportional to the number of lattice points taken. This is not altogether trivial since
nonperturbative metric fluctuations allow, in principle, ``crumpling'' of the space, and that is what some researchers
indeed typically observe in alternative nonperturbative approaches to gravity. In spinor gravity, it is a natural result
following from the non-compressibility of fermions. Second, the quantum average of the curvature turns out to be zero
such that the empty space without sources is effectively flat. This is also a welcome feature since the natural result
in nonperturbative gravity is that the curvature is of the order of the cutoff, that is of the Planck mass, which is
unacceptable. Third, despite flatness the theory definitely describes a fluctuating {\it quantum}
vacuum, as exemplified by the fact that the physical volume variance or susceptibility $\langle V^2\rangle\!-\!\langle V\rangle^2$
is nonzero.

As a result of the spontaneous breaking of continuous symmetry (here: chiral symmetry), a Goldstone field
appears. We check by an explicit calculation in Section VII that the low-momentum effective (``chiral'') Lagrangian
for the Goldstone field is diffeomorphism-invariant as expected. This invariance is rooted in the way we construct
the original lattice action for spinors. The appearance of a Goldstone particle means that a definite bilinear
combination of fermions is capable of propagating to large distances. However this is not enough: in order for
the system to totally loose memory about the original lattice, {\it all} degrees of freedom have to propagate
to long distances in lattice units. This happens only at a phase transition where we expect that the Einstein--Hilbert
action emerges as a low-energy effective action for the classical metric, with the cosmological constant being
automatically zero, see Section VIII. In Section IX we discuss the dimensions of various quantities and fields
used throughout the paper. We summarize in Section X.

\section{Composite frame fields}

Following Ref.~\cite{Diakonov:2011im} we introduce a composite frame field $e^A_\mu$ built as a bilinear
fermion ``current''. In $d$ dimensions the frame field transforms as a vector of the $SO(d)$ Lorentz gauge
group:
\begin{eqnarray}
e_\mu^A(x)\xrightarrow{\rm Lorentz} O^{AB}(x)e^B_\mu(x).
\la{e-transf-L}\end{eqnarray}
Since $A,B,\ldots = 1,\ldots, d$ are flat Euclidean indices we can equivalently write them either as subscripts
or superscripts. The frame field transforms also as rank-one tensor (world vector) with respect
to diffeomorphisms $x^\mu \to x'^\mu(x)$:
\begin{eqnarray}
e_\mu^A(x)\xrightarrow{\rm diffeomorphism} e^A_{\mu'}(x')\frac{\partial{x'}^{\mu'}}{\partial x^\mu}.
\la{e-transf-D}\end{eqnarray}

Our basic objects are fermion fields $\psi(x),\psi^\dagger(x)$ assumed to be world scalars under diffeomorphisms,
and transforming according to the spinor representation of the Lorentz group,
\bea\n
&&\psi(x)\to V(x)\psi(x),\qquad \psi^\dagger(x)\to \psi^\dagger(x)V^\dagger(x),\qquad V\in SO(d),\\
&&\psi,\psi^\dagger(x)\to\psi,\psi^\dagger\left(x'(x)\right).
\la{psi-transform}\eea
The dimension of the spinor representation is $d_f=2^{[d/2]}$, see, {\it e.g.}\cite{Zee:2003}.

We introduce the covariant derivative in the spinor representation,
\beq
\nabla_\mu=\partial_\mu-\frac{i}{2}\omega_\mu^{AB}\Sigma_{AB},\qquad
\overleftarrow{\nabla}_\mu=\overleftarrow{\partial}_\mu+\frac{i}{2}\omega_\mu^{AB}\Sigma_{AB}
\la{cov-der-spinor}\eeq
where $\omega^{AB}_\mu$ is the spin connection in the adjoint representation of the $SO(d)$ group, and
$\Sigma_{AB}$ are its $d_f\times d_f$ generators,
\beq
\Sigma_{AB}=\frac{i}{4}[\gamma_A\gamma_B],
\la{Sigma-def}\eeq
built from Dirac matrices $\gamma_A$ satisfying the Clifford algebra,
\beq
\{\gamma_A\gamma_B\}=2\delta_{AB}\,{\bf 1}_{d_f\times d_f}.
\la{Clifford}\eeq
In the adjoint (antisymmetric tensor) representation the corresponding covariant derivative is
\beq
D_\mu^{AB}=\partial_\mu \delta^{AB}+\omega_\mu^{AB}.
\la{cov-der-adjoint}\eeq
Its commutator defines the curvature
\beq
[D_\mu,D_\nu]^{AB}=\mathcal{F}^{AB}_{\mu\nu}, \qquad\[\nabla_\mu,\nabla_\nu\]=-\frac{i}{2}\mathcal{F}^{AB}_{\mu\nu}\Sigma_{AB},
\la{D-com}\eeq
where
\beq
{\cal F}^{AB}_{\mu\nu}=\partial_\mu\omega^{AB}_\nu+\omega^{AC}_\mu\omega^{CB}_\nu - (\mu\leftrightarrow\nu).
\la{F-def}\eeq

One can built two distinct bilinear combinations of the fermion fields, transforming as the frame field
\urs{e-transf-L}{e-transf-D}:
\bea
\la{e-def}
e^A_\mu &=& i(\psi^\dagger \gamma^A\nabla_\mu\psi+\psi^\dagger\overleftarrow{\nabla}_\mu \gamma^A\psi),\\
\la{f-def}
f^A_\mu &=& \psi^\dagger \gamma^A\nabla_\mu\psi-\psi^\dagger\overleftarrow{\nabla}_\mu \gamma^A\psi.
\eea
To check that $e^A_\mu$ and $f^A_\mu$ transform as a vector \ur{e-transf-L} one needs the relation between
the matrix $V$ rotating spinors \ur{psi-transform} and the matrix $O$ rotating vectors,
$$
O^{AB}=\frac{1}{d_f}\Tr\(V^\dagger \gamma^A V \gamma^B\).
$$
Given that $\psi,\psi^\dagger$ anticommute, the above bilinear operators are Hermitian.

We can define the bilinear fermion operator that plays the r\^ole of the torsion field, for example,
\beq
T^A_{\mu\nu}(e)\;\bydef\;\frac{1}{2}\left(D^{AB}_\mu e^B_\nu-D^{AB}_\nu e^B_\mu\right)
=\frac{i}{4}\mathcal{F}_{\mu\nu}^{AB}\left(\psi^\dagger\gamma_B\psi\right)
\la{torsion-def}\eeq
and similarly for the other composite frame field $f_\mu$ \ur{f-def}.

\section{Diffeomorphism-invariant action terms}

One can now construct a sequence of many-fermion actions that are invariant under local Lorentz transformations
and also diffeomorphism-invariant, using either $e^A_\mu$ or $f^A_\mu$ (or both) as building blocks:
\bea\la{actions}
S_k &=& \int\! d^dx\,\frac{1}{d!}\,\epsilon^{\mu_1\mu_2...\mu_d}\,\epsilon^{A_1A_2..A_d}\,
\left({\cal F}^{A_1A_2}_{\mu_1\mu_2}\ldots{\cal F}^{A_{2k-1}A_{2k}}_{\mu_{2k-1}\mu_{2k}}\right)\,
\left(e^{A_{2k+1}}_{\mu_{2k+1}}\ldots e^{A_d}_{\mu_d}\right)\,,\\
\n
k &=& 0,1,\ldots,[d/2],
\eea
where $\epsilon^{\mu_1\mu_2...\mu_d}$ is the totally antisymmetric (Levi-Civita) tensor. Notice that $S_0$ is the analog
of the cosmological term but there are many of them since one can replace any number of $e^A_\mu$'s by $f^A_\mu$'s,
$S_1$ is the analog of the Einstein--Hilbert--Cartan action linear in curvature, and the last action term $S_{[d/2]}$
for even $d$ is a full derivative. Apart from full derivatives, there are 3 possible action terms in $2d$, 6 terms in $3d$,
8 terms in $4d$, 12 terms in $5d$, {\it etc.}

The use of $\epsilon^{\mu_1\mu_2...\mu_d}$ is obligatory to support diffeomorphism-invariance. In principle,
one can construct Lorentz-invariant action terms by contracting the flat indices with Kronecker deltas instead
of $\epsilon^{A_1...A_d}$, however that will make the action term $P$- and $T$-odd. For example, there is
a well known $P,T$-odd term in four dimensions, called sometimes the Holst action,
$\epsilon^{\kappa\lambda\mu\nu}\,{\cal F}^{AB}_{\kappa\lambda}\,e^A_\mu e^B_\nu$~\cite{Hojman:1980kv,Nelson:1980ph},
but we do not consider such terms here.

One can add to the list of admissible action terms any of the actions \ur{actions} multiplied by
any power of the world and Lorentz-group scalar $(\psi^\dagger\psi)$; we shall consider
such kind of terms later on in relation to the spontaneous breaking of chiral symmetry.

All action terms \ur{actions} are apparently invariant under two global $U(1)$ rotations:
\begin{itemize}
\item phase rotation related to the fermion number conservation, $\psi \to e^{i\alpha}\psi,\;\psi^\dagger\to \psi^\dagger e^{-i\alpha}$
\item  chiral rotation for even dimensions $d$, $\psi \to e^{i\beta\gamma_{d+1}}\psi,\;\psi^\dagger\to \psi^\dagger e^{i\beta\gamma_{d+1}}$,
where $\gamma_{d+1}=i^{d/2}\gamma_1\gamma_2\ldots\gamma_d,\quad \{\gamma_{d+1}\gamma_d\}=0,\quad \gamma^2_{d+1}={\bf 1}$,
\end{itemize}
since both $e^A_\mu$ and $f^A_\mu$ are invariant under these transformations. The two corresponding N\"other currents are
conserved. However, both symmetries can be spontaneously broken by interactions, and we shall see that this is what indeed
typically happens.

\section{Spinor gravity on the lattice}

In order to formulate quantum theory properly one has to regularize it at short distances. The most clear-cut regularization
is by (lattice) discretization, however diffeomorphism-invariance imposes severe restrictions on it, see recent discussion
by Wetterich~\cite{Wetterich:2011yf,Wetterich:2011xv,Wetterich:2012qu}. We impose two basic requirements:
\begin{itemize}
\item explicit invariance under local gauge transformations of the Lorentz group, small or large
(as in lattice gauge theory)
\item if the fields vary slowly in lattice units, {\it i.e.} in the continuum limit, the lattice action reduces to
one of the diffeomorphism-invariant action terms \ur{actions} and the like.
\end{itemize}

\subsection{Triangulation by simplices}

To that end, we introduce an abstract discretized space where only the topology of vertices and edges connecting
neighbor vertices is chosen beforehand and fixed. We find that the simplest hypercubic topology does not work.
Only in two dimensions it is possible, for accidental reasons, to fulfill item 2 above by introducing a square
lattice. In higher dimensions, the simplest but sufficient construction is to use a simplicial lattice.
For uniformity, in two dimensions we also consider a triangle lattice made of three-vertex cells. In $3d$ simplices
are tetrahedra or 4-cells, in $4d$ these are pentachorons or 5-cells, and so on.

It is always possible to cover the whole $d$-dimensional space by $(d+1)$-cells or simplices, although
the number of edges entering one vertex may not be the same for all vertices. Alternatively, the number of
edges coming from all vertices is the same but then the edges lengths may vary, if one attempts to force
the lattice into flat space. Since only the topology of the nearest neighbors matters and the abstract
``number'' space does not need to be flat, this is also acceptable. The important thing is that the chosen
set of cells should fill in the space without holes and without overlapping~\cite{footnote-1}.

All vertices in a simplicial lattice can be characterized by a set of $d$ integers. For brevity
we label these $d$ numbers by a single integer $i$. Each vertex has its unique integer label
$i$, supplemented with a rule what labels are ascribed to the neighbor vertices forming elementary
cells. We shall denote the $d+1$ labels belonging to one cell by $i=0,1,\ldots,d$. In this section,
we write down the full lattice action as a sum over actions for individual simplicial cells,
therefore we shall not be concerned with the precise geometric arrangement of the cells.

Each vertex in the abstract number space corresponds to the real world coordinate by a certain
map $x^\mu(i)$. The goal is to write possible action terms in such a way that if the fields vary
slowly from one vertex (or link) to the topologically neighbor one, the action reduces to one of
the possible diffeomorphism- and local Lorentz-invariant action term in \Eq{actions}.

We start by writing the volume of an elementary cell (simplex) in a given coordinate system in $d$ dimensions.
It can be presented as a determinant of a $d\times d$ matrix,
\beq
V_{\rm simplex}=\frac{1}{d!}{\rm det}_{(\mu,i)} (x^\mu_i-x^\mu_0),
\la{V_simplex-1}\eeq
where $x_0^\mu$ is the coordinate ascribed to one of the vertices, and $x_i^\mu,\;i=1\ldots d$ are the
coordinates ascribed to all the other vertices. We introduce the notion of a ``positive order''
of vertices $i$ in the cell: it is such that for smooth functions $x^\mu_i$ the volume \ur{V_simplex-1} is positive.
An odd permutation of vertices in this set makes a ``negative order''.

It will be convenient to use the antisymmetric symbol
\beq
\epsilon^{i_0i_1...i_d}=\left\{
\begin{array}{cl}
0 ~~~& \text{if $i_k$ does not belong to a given cell,} \\
1 ~~~& \text{if the set $i_0,i_1...i_d$ is in the positive order,} \\
-1 ~~~& \text{if the set $i_0,i_1...i_d$ is in the negative order.}
\end{array}\right.
\la{epsilon}\eeq
With the help of this symbol the cell volume \ur{V_simplex-1} can be written as
\beq
V_{\rm simplex}=\frac{\epsilon^{i_0i_1i_2..i_d} }{(d+1)!}
\frac{\epsilon_{\mu_1\mu_2..\mu_d}}{d!}\left(x_{i_1}^{\mu_1}-x_{i_0}^{\mu_1}\right)
\left(x_{i_2}^{\mu_2}-x_{i_0}^{\mu_2}\right)\ldots\left(x_{i_d}^{\mu_d}-x_{i_0}^{\mu_d}\right).
\la{V_simplex-2}\eeq

\subsection{Lattice action}

The building blocks of our construction are anticommuting spinor fields $\psi_i,\psi^\dagger_i$
that are world scalars and ``live'' on lattice vertices $i$, and the parallel transporter $U_{ij}$.
As in any lattice gauge theory, we replace the connection $\omega_\mu$ by a unitary matrix
``living'' on lattice links~\cite{Diakonov:2011im},
\beq
U_{ij}=P\exp\(-\frac{i}{2}\int_{x_i}^{x_j}\omega^{AB}_\mu\Sigma^{AB} dx^\mu\),~~~~~~~~U_{ji}=U^\dagger_{ij}.
\la{U-def}\eeq
In terms of these lattice variables the discretized versions of the composite frame fields
\urs{e-def}{f-def} are:
\bea\la{tilde-e}
\tilde e_{i,j}^A &=& i(\psi^\dagger_j U_{ji}\gamma^A U_{ij}\psi_j-\psi^\dagger_i \gamma^A\psi_i),\\
\la{tilde-f}
\tilde f_{i,j}^A &=& \psi_i^\dagger \gamma^AU_{ij}\psi_j-\psi^\dagger_jU_{ji}\gamma^A\psi_i.
\eea
The difference between $\tilde e$ and $\tilde f$ is that the first has both fermions in the same vertex
whereas in the second fermions are residing in the neighbor vertices.

Expanding all fields in \Eqs{tilde-e}{tilde-f} around the center of a cell
$x=\frac{1}{d+1}\sum_{i=0}^d x_i$ we obtain:
\bea\label{e-cont}
\tilde e^A_{i,j} &=& (x_j^\mu-x_i^\mu)e_\mu^A(x)+\mathcal{O}(\Delta x^2), \\
\label{f-cont}
\tilde f^A_{i,j} &=& (x_j^\mu-x_i^\mu)f_\mu^A(x)+\mathcal{O}(\Delta x^2),
\eea
where $e^A_\mu,\,f^A_\mu$ are given by their continuum expressions \urs{e-def}{f-def}, and the
correction term is proportional to the derivatives of the fields and to the squares of the lengths
of the cell edges. If the fields are slowly varying, meaning that the derivatives are small, the
correction term can be neglected. This is what we mean by the continuum limit.

We also need the discretized version of the curvature tensor $\mathcal{F}_{\mu\nu}^{AB}$: it is a plaquette.
In our case the plaquettes are triangles, and we define the parallel transporter along a closed triangle
spanning the $i,j,k$ vertices:
\beq
P_{ijk}=U_{ij}U_{jk}U_{ki},~~~~~~~P^{AB}_{ijk}=\frac{1}{d_f}\Tr\(\Sigma^{AB}P_{ijk}\).
\la{W}\eeq
Expanding $P_{ijk}$ around the center of the cell $x$ we obtain:
$$
P_{ijk}=1-\frac{i}{4}(x^\mu_j-x^\mu_i)(x_k^\nu-x_i^\nu)\mathcal{F}_{\mu\nu}(x)+\mathcal{O}(\Delta x^3)
$$
and
$$
P^{AB}_{ijk}=-\frac{i}{4}(x^\mu_j-x^\mu_i)(x_k^\nu-x_i^\nu)\mathcal{F}^{AB}_{\mu\nu}(x)+\mathcal{O}(\Delta x^3).
$$

Using the above ingredients one can easily construct the lattice regularized version of the action terms \ur{actions}.
For example, the discretized cosmological term $S_0$ has the form:
\beq
\tilde S_0=\sum_{\rm all\;cells} \frac{\epsilon^{i_0i_1\ldots i_d}}{(d+1)!}\,
\frac{\epsilon^{A_1A_2..A_d}}{d!}\,\tilde e^{A_1}_{i_0i_1}\tilde e^{A_2}_{i_0i_2}..\tilde e^{A_d}_{i_0i_d}
\la{tildeS-def}\eeq
where any number of $\tilde e$'s can be replaced by $\tilde f$'s. In the continuum limit one uses \Eqs{e-cont}{f-cont}
and obtains
\beq
\tilde S_0=\sum_{\rm all\;cells} \frac{\epsilon^{i_0i_1..i_d}}{(d+1)!}\,\frac{\epsilon^{\mu_1\mu_2..\mu_d}}{d!}\,
\left(x_{i_1}^{\mu_1}-x_{i_0}^{\mu_1}\right)\ldots \left(x_{i_d}^{\mu_d}-x_{i_1}^{\mu_d}\right)\,
\det(e)\[1+\mathcal{O}(\Delta x)\].
\la{tildeS}\eeq
The coordinate factors combine into the volume of the cell \ur{V_simplex-2} and one gets
\beq
\tilde S_0=\sum_{\rm cells} V({\rm cell})\det(e)\[1+\mathcal{O}(\Delta x)\] \rightarrow \int d^d x\det(e)=S_0,
\la{S01}\eeq
where $\det(e)$ is composed from the continuum tetrad \ur{e-def} and is attributed to the center of a cell.
\Eq{S01} proves that the lattice action \ur{tildeS-def} becomes the needed continuum action \ur{actions} if the fields
involved are slowly varying from one lattice vertex to the neighbor ones.

Similarly, one finds the lattice version of all other action terms $S_k$ of \Eq{actions}:
\beq
\tilde S_k=(4i)^k\sum_{\rm cells} \frac{\epsilon^{i_0i_1\ldots i_d}}{(d+1)!}\,\frac{\epsilon^{A_1A_2\ldots A_d}}{d!}\,
\left(P^{A_1A_2}_{i_0i_1i_2}\,P^{A_3A_4}_{i_0i_3i_4}\ldots P^{A_{2k-1}A_{2k}}_{i_0i_{2k-1}i_{2k}}\right)
\left(\tilde e^{A_{2k+1}}_{i_0i_{2k+1}}\ldots\tilde e^{A_d}_{i_0i_d}\right)\quad\rightarrow\quad S_k,
\la{tildeSk}\eeq
where the total number of plaquette factors $P$ \ur{W} is $k$, $k=0,1\ldots[d/2]$. In fact one can write
a variety of such action terms replacing any number of composite frame fields $\tilde e$ \ur{tilde-e}
by the composite frame fields $\tilde f$ \ur{tilde-f}.

\subsection{Lattice partition function}

The lattice-regularized partition function for the spinor quantum gravity is quite similar to that
of the common lattice gauge theory. One integrates with the Haar measure over link variables $U_{ij}$
living on lattice edges, and over anticommuting fermion variables $\psi_i,\psi^\dagger_i$ living
on lattice sites. The lattice, though, must be simplicial, otherwise the trick used {\it e.g.} in
\Eq{tildeS} to get the diffeomorphism-invariant action in the continuum limit, would not work.

Because of the requirement of diffeomorphism-invariance, the lattice action is quite different
from those used in common lattice gauge theory. Typically one has many-fermion terms in the action.
There are no action terms without fermions. One can write 3 action terms in $2d$ (all of them are 4-fermion),
6 terms in $3d$ (four are 6-fermion and two are 2-fermion), 8 terms in $4d$ (five are 8-fermion and three
are 4-fermion), {\it etc.} We assume that spinor fields are dimensionless since we normalize the basic
Berezin integrals as
\beq
\int d\psi\,\psi =1,\qquad \int d\psi^\dagger\,\psi^\dagger =1,\qquad \int d\psi=0,\qquad \int d\psi^\dagger=0,
\la{Berezin}\eeq
hence all quantities in \Eq{tildeSk} are dimensionless. Therefore, the ``coupling constants'' $\lambda_k$
one puts as arbitrary coefficients in front of the action terms $\tilde S_k$ \ur{tildeSk} are all dimensionless.

The partition function is
\beq
{\cal Z}=\prod_{{\rm vertices}\; i}\int d\psi^\dagger_i\,d\psi_i\;\prod_{{\rm links}\;ij}\int dU_{ij}\,
\exp\left(\sum_{\rm cells}\lambda_k^{(m)}\tilde S_k^{(m)}(\psi^\dagger,\psi, U)\right)
\la{Z}\eeq
where $\tilde S_k^{(m)}$ are lattice actions of the type \ur{tildeSk} with any number of composite
frame fields $\tilde e$ \ur{tilde-e} replaced by the other composite frame fields $\tilde f$ \ur{tilde-f}.

\section{Mean-field approximation}

The partition function \ur{Z} defines a new type of a theory, and new methods -- exact, numerical
and approximate -- have to be developed.

In principle, in order to compute the partition function \ur{Z} as well as correlation functions,
{\it etc.}, one has to Taylor-expand the exponent in \Eq{Z} to certain powers of the fermionic action
terms $S_k$ such that at all lattice sites there is precisely the same number of fermion operators
$\psi^\dagger$ and $\psi$ as there are integrations, since all other contributions are identically
zero by the Berezin integration rule \ur{Berezin} for anticommuting variables. The subsequent
integration over link variables with the Haar measure is simple~\cite{Diakonov:2011im} since
link matrices $U_{ij}$ never appear in a large power. Moreover, the majority of potentially possible
contributions are killed by link integration.

In practice, however, the arising combinatorial problem is tremendous, and we did not manage yet
to find a computational algorithm that would be faster than the exponent of the lattice volume.
So far we have done a toy model in $1d$ exactly (see the Appendix) and succeeded in computing
numerically correlation functions in $2d$ for limited volumes. There is a hope that the $2d$ model
may be solved exactly but the method can hardly be extended to higher dimensions.

Therefore, for this pilot study, we have developed an approximate mean-field method to get the first
glance on the dynamics of the new interesting theory at hand. Comparing the results with an exactly
solvable model we see that the mean field accuracy is within a few percent. In the $2d$ model there
are a few exact functional relations that are satisfied with the accuracy better than 15\%, and this
can be systematically improved. More important, the mean field approximation reveals a nontrivial phase
structure of the theory in the space of the coupling constants $\lambda_k$. This is the main finding
of this study that may have important physical implications, see the Introduction.

The mean-field approximation we use is an extension of methods developed in condensed matter physics, that
go under the name of ``dynamical mean-field approach'' or ``local impurity self-consistent approximation''
or ``cavity method'', see Ref.~\cite{Georges:1996zz} for a review. Roughly speaking, the idea of
the method is the following: One first picks up a simple element of the lattice ({\it e.g.}
one simplex or a group of simplices with or without the boundary, let us call this fixed element
``the cavity''), and calculates the effective action for the fields inside the cavity in the collective
background of the fields outside it, replacing the background by the supposed mean field. At the second stage,
one makes the method self-consistent, namely one calculates the mean field by integrating over the
``live'' variables inside the chosen cavity using the effective action found and expressed through
the mean field at the first stage. As a result one gets a system of highly nonlinear self-consistent
equations for a set of mean values of the field operators. Solving those equations one obtains
the mean-field values as function of the coupling constants $\lambda_k^{(m)}$. This gives the
phase diagram of the theory in the space of the coupling constants.

The method has the advantage that it can be systematically improved by enlarging the chosen cavity.
In the limit when the cavity covers the whole lattice it is an exact calculation. Also, it is known
that the accuracy of the mean field method is better the more nearest neighbors there are~\cite{Georges:1996zz}.
In simplicial lattices the number of neighbor cells is large, and the mean field method becomes
exact in the limit $d\to\infty$.

Let us formulate the method more mathematically. We choose the cavity, for example the elementary simplex
with the boundary. We label the ``live'' fields belonging to the cavity by $m,n,\ldots$ and the fields outside
the cavity (that will be replaced by mean fields) by $i,j,\ldots$. The full partition function can be written
symbolically as
\beq
{\cal Z}=\int d\psi^\dagger_m d\psi_m dU_{mn}\,e^{S_{mn}}\;\int dU_{mi}\,e^{S_{mi}}\int d\psi^\dagger_i
d\psi_i dU_{ij}\,e^{S_{ij}}
\la{Z1}\eeq
where $S_{mn}$ is the part of the action that contains only fields from the cavity, $S_{ij}$ contains
only fields from outside the cavity and $S_{mi}$ contains both. The link elements $U_{mi}$ are connecting
vertices from the cavity with their nearest neighbors outside.

The last integral in \Eq{Z1} is the full partition function with the cavity cut out. When the lattice
volume goes to infinity cutting out a finite cell does not change the averages of operators as compared
to the averages computed on a full lattice; we denote them as
\beq
\langle\langle O\rangle\rangle = \frac{\int d\psi^\dagger_i d\psi_i dU_{ij}\,e^{S_{ij}}\,O(\psi^\dagger_i,\psi_i, U_{ij})}
{\int d\psi^\dagger_i d\psi_i dU_{ij}\,e^{S_{ij}}}.
\la{avO1}\eeq

The integration over the links $U_{mi}$ connecting the cavity with the outside neighborhood
must be performed explicitly in \Eq{Z1}. We expand $e^{S_{mi}}$ in powers of the mixed action;
since $S_{mi}$ is a fermion operator the power series is finite. Integrating over $U_{mi}$
splits all terms involved into a sum of products of operators composed of the cavity fields
$O\left(\psi^\dagger_m, \psi_m, U_{mn}\right)$ and those living outside the cavity
$O'\left(\psi^\dagger_i, \psi_i, U_{ij}\right)$:
\beq
\int dU_{mi}\,e^{S_{mi}}= 1+\sum_p O_p\left(\psi^\dagger_m, \psi_m, U_{mn}\right)\,
O'_p\left(\psi^\dagger_i, \psi_i, U_{ij}\right)
\la{Z2}\eeq
where the sum goes over various fermion operators labeled by $p$. Operators built from the
cavity fields are left intact whereas the outside operators are replaced by the averages
according to \Eq{avO1}. We, thus, obtain the effective action for the fields inside the cavity:
\beq
e^{S_{{\rm eff}, mn}} = e^{S_{mn}}\,
\left(1+\sum_p O_p\left(\psi^\dagger_m, \psi_m, U_{mn}\right)\,\langle\langle O'_p \rangle\rangle\right).
\la{Seff}\eeq
Finally, we make the calculation self-consistent by requesting that the operator averages $\langle O_p\rangle$
computed from the cavity fields alone with the effective action \ur{Seff} coincide with the full ones
$\langle\langle O_p \rangle\rangle$:
\beq
\langle O_p\rangle
=\frac{\int d\psi^\dagger_m d\psi_m dU_{mn}\,e^{S_{{\rm eff},mn}}O_p\left(\psi^\dagger_m, \psi_m, U_{mn}\right)}
{\int d\psi^\dagger_m d\psi_m dU_{mn}\,e^{S_{{\rm eff},mn}}}\quad = \quad \langle\langle O_p \rangle\rangle.
\la{self-cons}\eeq
Since $S_{\rm eff}$ depends on the averages $\langle\langle O_p \rangle\rangle$ the self-consistency \Eq{self-cons}
is in fact a set of nonlinear equations on the mean values of the operators introduced in this derivation. Solving
those equations one finds the values of the average operators as function of the coupling constants of the theory.

Of special interest are the cases where certain operator averages (the ``condensates'') violate the continuous
symmetries of the original theory. It signals the spontaneous breaking of symmetry and leads to a nontrivial
phase diagram for the theory. In the next section we illustrate it in a general $2d$ model.

\section{Two-dimensional spinor gravity}

The partition function is defined by \Eq{Z} where the action has in general three terms
with three arbitrary coupling constants $\lambda_{1,2,3}$,
\beq
S = \int d^2x\left(\lambda_1 \det(e)+\lambda_2 \det(f)
+\lambda_3 \frac{1}{2!}\,\epsilon^{AB}\,\epsilon^{\mu\nu}\,e^A_\mu f^B_\nu\right),\qquad A,B=1,2.
\la{S0}\eeq
The lattice-regularized version of it is, according to \Eq{tildeS-def},
\beq
\tilde S =\sum_{\rm cells} \frac{\epsilon^{ijk}}{3!}\frac{\epsilon^{AB}}{2!}\,
\(\lambda_1\,\tilde e^A_{ij}\tilde e^B_{ik}+\lambda_2\, \tilde f^A_{ij}\tilde f^B_{ik}
+\lambda_3\,\tilde e^A_{ij}\tilde f^B_{ik}\),
\la{tildeS0}\eeq
where $i,j,k=0,1,2$ label the vertices of a cell which in $2d$ is a triangle.
Using integration by parts the first term in \ur{S0} can be rewritten as
$-\mathcal{F}_{12}^{12} (\psi^\dagger\psi)^2$. It gives an alternative discretization for the same continuum action:
\beq
\tilde S =\sum_{\rm cells} \frac{\epsilon^{ijk}}{3!}\frac{\epsilon^{AB}}{2!}\,
\(-\frac{i}{3}\lambda_1\,P^{AB}_{ijk}(\psi^\dagger_i\psi_i)^2+\lambda_2\, \tilde f^A_{ij}\tilde f^B_{ik}
+\lambda_3\,\tilde e^A_{ij}\tilde f^B_{ik}\).
\la{tildeS0-number2}\eeq
Although in the continuum limit the lattice actions (\ref{tildeS0}) and (\ref{tildeS0-number2}) differ by a full
derivative, the lattice mean-field approximation gives numerically slightly different results depending on whether
we start from \Eq{tildeS0} or from \Eq{tildeS0-number2}. The deviation serves as one of the checks of the accuracy
of the approximation, and we find it consistent with other accuracy checks.

In $d=2$ the Lorentz group is the Abelian $SO(2)\simeq U(1)$ group. The spinors are two-component,
and the $\gamma$-matrices are the Pauli matrices $\gamma^A=\sigma^A$, $A=1,2$. The Lorentz rotations
generator is $\Sigma^{12}=-\sigma^3/2$, see \Eq{Sigma-def}. The analog of the ``gamma-five'' matrix
in $2d$ is $\gamma^3=i\gamma^1\gamma^2=-\sigma^3$.

Both variants of the frame field $e^A_\mu$ and $f^A_\mu$ as well as there lattice extensions,
$\tilde e^A_{ij}$ and $\tilde f^A_{ij}$, are invariant under two global $U(1)_V\times U(1)_A$ transformations:
\bea\la{vector}
{\rm vector\;transformation:}&&\qquad\psi\rightarrow e^{i\frac{\beta}{2}}\psi,\qquad
\psi^\dagger\rightarrow \psi^\dagger e^{-i\frac{\beta}{2}},\\
\la{axial}
{\rm axial\;transformation:}&&\qquad\psi\rightarrow e^{i\frac{\alpha}{2} \sigma^3}\psi,\qquad
\psi^\dagger\rightarrow \psi^\dagger e^{i\frac{\alpha}{2} \sigma^3}.
\eea
Therefore, both the continuum \ur{S0} and the lattice \ur{tildeS0} actions possess these two global
symmetries also; the corresponding N\"other currents are conserved. The vector symmetry means that
the fermion number is conserved whereas the axial means that the difference between the numbers
of ``left-handed'' and ``right-handed'' fermions (described by the upper and lower components of
the spinors, respectively) is also conserved. It is also called the helicity conservation,
or chiral symmetry.

\subsection{Exact results}

In the $2d$ partition function \ur{Z} there are four integrals per site over fermion variables
$\psi^1,\psi^2,\psi^\dagger_1,\psi^\dagger_2$, and one integration per link over the Abelian
matrix $U_{ij}=\exp\left(-i\frac{\omega_{ij}}{4}\sigma^3\right)$. Berezin's integrals over fermions
\ur{Berezin} are non-zero only when every lattice site takes exactly four fermion operators
from the action exponent. Meanwhile, each term in the action \ur{tildeS0} or \ur{tildeS0-number2}
is four-fermion. From counting the number of fermion fields coming from the action (which must
be equal to the number of integrations) we conclude that the partition function $Z$ is a homogenous
polynomial of the coupling constants $\lambda_{1,2,3}$ of order $N$, where $N$ is the total number
of sites in the lattice,
\beq
{\cal Z} = \lambda_1^N\,F\left(\frac{\lambda_2}{\lambda_1},\frac{\lambda_3}{\lambda_1}\right)
\rightarrow \left\{\begin{array}{cc}
C_1\lambda_1^N, & \quad\lambda_{2,3}\to 0, \\
C_2\lambda_2^N, & \quad\lambda_{1,3}\to 0, \\
C_3\lambda_3^N, & \quad\lambda_{1,2}\to 0.\end{array}\right.
\la{Z-lambda}\eeq
Since there are two types of frame fields, $e$ and $f$, we can define three types of ``physical''
or invariant volumes of the generally curved space, averaged over quantum fluctuations of the
composite frame fields,
\bea\la{V1}
\langle V_1\rangle &\bydef & \langle \int d^2x \det(e)\rangle
=\frac{1}{{\cal Z}}\,\frac{\partial {\cal Z}}{\partial \lambda_1}
= \frac{\partial\log{\cal Z}}{\partial \lambda_1},\\
\la{V2}
\langle V_2\rangle &\bydef & \langle \int d^2x \det(f)\rangle
= \frac{1}{{\cal Z}}\,\frac{\partial {\cal Z}}{\partial \lambda_2}
= \frac{\partial\log{\cal Z}}{\partial \lambda_2},\\
\la{V3}
\langle V_3\rangle &\bydef & \langle \int d^2x\,\frac{1}{2!}\,\epsilon^{AB}\,\epsilon^{\mu\nu}\,e^A_\mu f^B_\nu \rangle
= \frac{1}{{\cal Z}}\,\frac{\partial {\cal Z}}{\partial \lambda_3}
= \frac{\partial\log{\cal Z}}{\partial \lambda_3}.
\eea
The immediate conclusion from \Eqsss{Z-lambda}{V3} is that the average action is
\beq
\langle S\rangle=\lambda_1\langle V_1\rangle + \lambda_2\langle V_2\rangle + \lambda_3\langle V_3\rangle
= N = \frac{M}{2}
\la{<S>}\eeq
irrespectively of the coupling constants, where $M$ is the number of triangle cells, which is
twice the number of vertices $N$ for large simplicial lattices.

Further on, one can introduce ``physical volume susceptibility'' or variance
\bea\la{DeltaV1}
\langle \Delta V_1^2\rangle &\bydef & \langle (V_1-\langle V_1\rangle)^2\rangle
= \langle V_1^2\rangle - \langle V_1\rangle^2 = \frac{\partial^2\log{\cal Z}}{\partial \lambda_1^2},\\
\la{DeltaV2}
\langle \Delta V_2^2\rangle &\bydef & \langle (V_2-\langle V_2\rangle)^2\rangle
= \langle V_2^2\rangle - \langle V_2\rangle^2 = \frac{\partial^2\log{\cal Z}}{\partial \lambda_2^2},\\
\la{DeltaV3}
\langle \Delta V_3^2\rangle &\bydef & \langle (V_3-\langle V_3\rangle)^2\rangle
= \langle V_3^2\rangle - \langle V_3\rangle^2 = \frac{\partial^2\log{\cal Z}}{\partial \lambda_3^2}.
\eea
Therefore from \Eq{Z-lambda} we know exactly the average physical volumes and volume susceptibilities
at least at the edges of the parameter space $\Lambda\bydef(\lambda_1,\lambda_2,\lambda_3)$:
\bea\la{V-edge}
&&\langle V_1\rangle_{\lambda_{2,3}\to 0}\quad = \quad \frac{M}{2\lambda_1},\qquad
\langle V_2\rangle_{\lambda_{1,3}\to 0} \quad = \quad \frac{M}{2\lambda_2},\qquad
\langle V_3\rangle_{\lambda_{1,2}\to 0} \quad = \quad \frac{M}{2\lambda_3},\\
\la{DeltaV-edge}
&&\langle \Delta V_1^2\rangle_{\lambda_{2,3}\to 0}\; = \; -\frac{M}{2\lambda_1^2},\qquad
\langle \Delta V_2^2\rangle_{\lambda_{1,3}\to 0}\; = \; -\frac{M}{2\lambda_2^2},\qquad
\langle \Delta V_3^2\rangle_{\lambda_{1,2}\to 0}\; = \; -\frac{M}{2\lambda_3^2},
\eea
where $M=2N$ is the total number of simplicial cells in the lattice. The proportionality
of these quantities to $M$ is a very general property (valid not only at the edges of the parameter
space but everywhere) following from \Eq{Z-lambda}. It shows that the physical volume is
an extensive quantity, as it should be. This is not altogether trivial since nonperturbative
metric fluctuations allow, in principle, ``crumpling'' of the space, or the formation of
``branched polymers'', and that is what some researchers observe in alternative nonperturbative
approaches to gravity. In spinor gravity, it is a natural result following physically from
the non-compressibility of fermions and mathematically expressed by \Eq{Z-lambda}.

The susceptibilities \ur{DeltaV-edge} are also extensive, as should be expected. In the classical
ground state there are no quantum fluctuations, so $\Delta V = 0$. The fact that \ur{DeltaV-edge} is
nonzero means that we are dealing with a fluctuating quantum vacuum. At the same time for large
volumes the relative strength of the fluctuations die out: $\sqrt{\Delta V^2}/V \sim 1/\sqrt{M}\to 0$.

There are theorems for mixed derivatives, valid in the whole parameter space
$\Lambda$, that can be used to check the accuracy of approximate calculations, for example,
\beq
\frac{\partial\; \langle \int d^2x \det(e)\rangle}{\partial\lambda_2}
=\frac{\partial\log{\cal Z}}{\partial\lambda_1\partial\lambda_2}
=\frac{\partial\; \langle \int d^2x \det(f)\rangle}{\partial\lambda_1}\,.
\la{mixed}\eeq

Finally, there is an exact statement about the average curvature. The number of link variables
in all terms of the action \ur{tildeS0} is even. That gives a nonzero result from integration
over links for the partition function. However, if one attempts to compute the average of
the curvature proportional to ${\cal F}$ that in the lattice formulation is given by a product
of three links (see \Eq{W}), the number of link variables becomes odd, and link integration
yields an identical zero. Therefore, we conclude that the average Cartan curvature proportional
to the average scalar curvature is zero,
\beq
\langle \det(e) R\rangle =2\langle {\cal F}^{12}_{12}\rangle =0.
\la{Rzero}\eeq
This result in $2d$ is, of course, in conformity with the zero Euler characteristic of a torus;
no other result could be correct. It is illuminating, however, to see how ``microscopically''
the Euler theorem works for fluctuating spaces. In higher dimensions $\det(e)R$ is not a full
derivative but it still may be possible to find its average in a similar way.

At the same time, owing to quantum fluctuations the average curvature squared is generally nonzero
and extensive,
\beq
\langle \left(\int d^2x \det(e) R\right)^2\rangle\quad\sim\quad M,
\la{RR}\eeq
implying that the volume-independent combination dies out in the thermodynamic limit,
\beq
\frac{\sqrt{\langle \left(\int d^2x \det(e) R\right)^2\rangle}}
{\langle\int d^2x \det(e)\rangle}\quad\sim\quad \frac{1}{\sqrt{M}}\to 0.
\la{sqrtRR}\eeq
\Eqss{V-edge}{Rzero}{sqrtRR} mean that although we apparently deal with a quantum fluctuating
vacuum, the space is on the average large and flat in the absence of external sources. Therefore,
one can say that the model describes a flat background metric $G_{\mu\nu}$ that is a unity matrix
in a particular frame representing the flat space but transforms as a tensor under the change
of coordinates. We shall use this notion in Section VII.

\subsection{Mean-field approximation for one simplex cavity}

In this Subsection we apply the mean-field method formulated in Section V to the lattice action
\ur{tildeS0-number2} where we first put for simplicity $\lambda_3=0$. At the end of this Section
we formulate the main results for $\lambda_3\neq 0$.

In the first approximation to the mean-field method, we choose the elementary triangle cell $(m,n,p)$
as the ``cavity'', see Fig.~3. The fields inside the triangle cavity are considered as real quantum
fields, whereas the fields outside the cavity are combined into certain gauge-invariant operators that
are frozen to their mean-field values. The triangle cavity is surrounded by three ``black'' triangles
of the type $(i,m,n)$ with a common edge, and by nine ``white'' triangles of the type $(i,j,m)$
with a common vertex.  The effective action for the fields inside the cavity gets contributions
from both types of neighbors.

\begin{figure}[thb]
\includegraphics[width=0.4\textwidth]{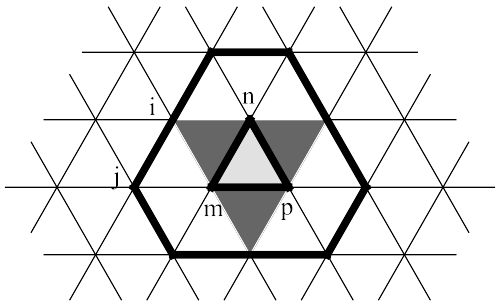}
\caption{The simplest triangle cavity $(m,n,p)$ and its neighbors used in the mean field calculation.}
\end{figure}

Following the method of Section V, we expand the action exponent for every border cell,
and integrate over the link variables $U_{mi}$ connecting the cavity with the outer lattice.
As the result we obtain the product of operators built of the fields inside the cavity and those built
of the outside fields. The latter operators are replaced by the averages to be found later
from the self-consistency condition. We stress that after integrating over $U_{im}$ the
operators on both sides can be only gauge-invariant.

For the single cell cavity we obtain operators of two types: single-site operators (they arise
from the ``white'' cells), and double-site operators built from fermions at adjacent vertices
(they arise from ``black'' cells).

Here is the list of operators that appear in this calculation. First of all, there are operators
that are invariant under the $U(1)_V\times U(1)_A$ transformations described in Subsection A:
\bea\n
O_1(i)&=&(\psi^\dagger_i\psi_i)^2, \\
\n
O_2(i,j)&=&(\psi^\dagger_iU_{ij}\psi_j)^2+(\psi^\dagger_jU_{ji}\psi_i)^2
+(\psi_i^\dagger\psi_i)(\psi_j^\dagger\psi_j)
-(\psi_i^\dagger\sigma^3\psi_i)(\psi_j^\dagger\sigma^3\psi_j),\\
O_3(i,j)&=&(\psi^\dagger_i\psi_i)^2(\psi^\dagger_j\psi_j)^2.
\la{O123}\eea
$O_1$ is a single-site operator while $O_2$ and $O_3$ are double-site operators.

To be able to study the potential breaking of the $U(1)_V\times U(1)_A$ symmetries we introduce
operators that transform under those rotations. The chiral non-invariant operators that transform
under $U(1)_A$ \ur{axial} are
\bea\n
C_1(i)&=&i(\psi^\dagger_i\psi_i),\qquad\overline C_1(i)=i(\psi^\dagger_i\sigma^3\psi_i,)\\
\n
C_2(i,j)&=&i\[(\psi^\dagger_i\psi_i)^2(\psi^\dagger_j\psi_j)
+(\psi^\dagger_j\psi_j)^2(\psi^\dagger_i\psi_i)\],\\
\overline  C_2(i,j)&=&i\[(\psi^\dagger_i\psi_i)^2(\psi^\dagger_j\sigma^3\psi_j)
+(\psi^\dagger_j\psi_j)^2(\psi^\dagger_i\sigma^3\psi_i)\].
\la{C12}\eea
Fermion number violating operators transforming under $U(1)_V$ \ur{vector} are
\bea\n
W_1(i)&=&\psi_{i,1}\psi_{i,2},\\
W_2(i,j)&=&(\psi^\dagger_i\psi_i)^2\psi_j^1\psi_j^2+(\psi^\dagger_j\psi_j)^2\psi_i^1\psi_i^2.
\la{W12}\eea
All operators are Hermitian.

The effective action for the fields inside the triangle cavity is computed as described in Section V.
From the ``black'' neighbors we obtain the double-site effective action
\bea\nn
e^{S(m,n)}&=&
1+\frac{8}{9}\lambda_1^2\[\(O_1(m)+O_1(n)\)\langle O_1\rangle + O_3\]\\
& + &\frac{\lambda_2^2}{36}\[O_2 \langle O_1\rangle
-2\( W_2^\dagger\langle W_1\rangle+ W_2\langle W_1^\dagger\rangle\)
- C_2\langle C_1\rangle+ \overline C_2\langle \overline C_1\rangle \]\,
\la{S-black2}\eea
where all double-site operators refer to the cavity vertices $m$ and $n$.
From the ``white'' neighbors we obtain the one-site effective action
\bea\nn
e^{S(m)}&=&
1+\frac{8}{9}\lambda_1^2\[2O_1 \langle O_1\rangle+\langle O_3\rangle\]\\
& + &\frac{\lambda_2^2}{36}\[ O_1\langle O_2\rangle
-2\( W_1^\dagger\langle W_2\rangle+ W_1\langle W_2^\dagger\rangle\)
- C_1\langle C_2\rangle + \overline C_1\langle\overline C_2\rangle \].
\la{S-white2}\eea
Actually the operator averages in \urs{S-black2}{S-white2} imply averaging over the cyclic
permutation of lattice sites in the cavity: {\it e.g.}
$\langle O_1\rangle=\frac{1}{3}\langle O_1(m)+O_1(n)+O_1(p)\rangle$,
and similarly for the double-site operators. The full effective action for
the cavity is a sum over all 12 neighbor cells,
\beq
e^{S_{\rm eff}}=\exp\left[S(m,n,p)+\left(S(m,n)+S(n,p)+S(p,m)\right)
+3\left(S(m)+S(n)+S(p)\right)\right],
\la{Seff-2D}\eeq
where $S(m,n,p)$ is the original action for the cavity triangle $(m,n,p)$, as given by \Eq{tildeS0-number2}.

We see that the effective action for the fields living in the cavity cell depend explicitly on the
yet unknown operator averages $\langle O\rangle,\,\langle C\rangle,\,\langle W\rangle$. To find them,
one equates the operator averages as defined by the effective action \ur{Seff-2D} to those introduced
previously, see \Eq{self-cons}. As the result one obtains a system of nonlinear self-consistency
equations on the averages $\langle O\rangle,\,\langle C\rangle,\,\langle W\rangle$. Solving those
equations one finds the averages as function of the coupling constants $\lambda_{1,2}$.

This calculation is straightforward but the equations are rather lengthy. Therefore, we
just comment here on its most important features.

First of all, we notice that $S_{\rm eff}$ is quadratic in the symmetry breaking operators $C_{1,2}$
and $W_{1,2}$, as it should be, therefore one gets a system of linear homogeneous self-consistency equations
on the averages $\langle C_{1,2}\rangle,\,\langle W_{1,2}\rangle$ that always have a zero solution,
unless the determinant of this set of linear equations is zero. If the determinant is nonzero
in the whole range of the parameter space $\Lambda$ there is no spontaneous symmetry breaking.
If the determinant passes through zero at some surface in the $\Lambda$ space, it is where
the second order phase transition takes place. Inside the domain where one of the $U(1)$ symmetries
is spontaneously broken the condensates $\langle C_{1,2}\rangle$ or $\langle W_{1,2}\rangle$
are nonzero and are found as anomalous solutions of the nonlinear equations, together with the
symmetry-preserving averages $\langle O_{1,2,3}\rangle$.

In the absence of symmetry breaking one puts $\langle C_{1,2}\rangle=\langle W_{1,2}\rangle =0$
and solves the system of three nonlinear equations on the averages $\langle O_{1,2,3}\rangle$.
There are in general several solutions but none are real in the whole $\Lambda$ space.
We pick up the solution that is real near the line $\lambda_2=0$. However it develops a
cut and becomes complex at the lines $|\lambda_2|=8.69 |\lambda_1|$ signalling that there
can be a phase transition along these lines. A careful study in the next Subsection shows that
indeed these are the border lines separating the phase with spontaneous chiral symmetry breaking,
see Fig.~4.

\begin{figure}[h]
\includegraphics[width=0.4\textwidth]{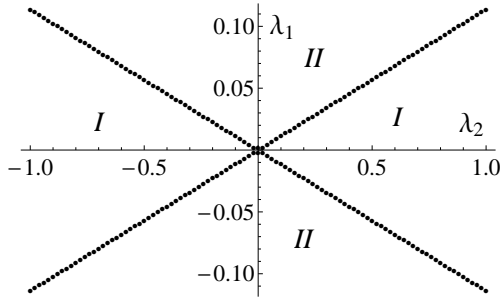}
\caption{The phase diagram of the $2d$ spinor gravity in the ($\lambda_1$,$\lambda_2$) plane at $\lambda_3=0$.
Region $I$ corresponds to the chiral-symmetry broken phase, region $II$ is a regular phase.
The dots show the lines of the 2$^{\rm nd}$ order phase transition:
$|\lambda_2|\simeq 8.69 |\lambda_1|$.}
\end{figure}

Outside this domain, {\it i.e.} at $|\lambda_2|<8.69 |\lambda_1|$ chiral symmetry is not
broken, the solution for the normal, symmetry-preserving operators is real, and one can approach
the line $\lambda_2=0$ where we can check the accuracy of the mean field method by comparing
the average physical volume $\langle V_1\rangle \approx\sum_{\rm cells}\langle\det(\tilde e)\rangle$
where the average is computed over one (cavity) cell with the effective action \ur{Seff-2D}, with the
exact result \ur{V-edge}. We find numerically
\beq
\langle V_1\rangle_{\lambda_{2,3}\to 0} = 0.572\,\frac{M}{\lambda_1}\quad({\rm mean\; field})
\qquad vs.\qquad 0.5\,\frac{M}{\lambda_1}\quad({\rm exact}),
\la{acc1}\eeq
where $M$ is the total number of lattice cells. We note that the functional dependence on
$\lambda_1$ is correct whereas the numerical coefficient deviates from the exact one by 15\%. A more
powerful check comes from computing the average action $\langle S\rangle $ which turns out to be a constant
up to the third digit in the whole range of analyticity in $\lambda_{1,2,3}$, equal to 0.57, instead
of the exact result \ur{<S>} being 0.5. This is the typical accuracy with which other checks with exact
results are fulfilled.

We have also tested a more primitive mean-field approximation where the cavity is taken in the form
of two neighbor vertices connected by a link. It is also capable of detecting the spontaneous breaking
of chiral symmetry but the accuracy is, of course, worse: it is at level of 40\%.

\subsection{Spontaneous chiral symmetry breaking}

An accurate way to study spontaneous breaking of a continuous symmetry is to introduce a small
term in the action that violates the symmetry in question explicitly. Since we are interested in
the spontaneous breaking of the $U(1)_A$ or chiral symmetry we introduce the simplest
diffeomorphism-invariant ``mass term''
\beq
S_{\chi{\rm -odd}}=\int d^2x \det(\tilde e)\,i m \,\psi^\dagger\psi
\la{mass-term}\eeq
that is not invariant under chiral rotations \ur{axial}. Its discretized lattice version is obvious,
see the first term in \Eq{tildeS0}.

Adding this term we repeat the same mean-field derivation of the effective action for the
triangle cavity as in \Eqs{S-black2}{S-white2} which now obtain an addition
\bea\la{Zeff-m1}
S_{\chi{\rm-odd}}(m,n)&=&\frac{\lambda_2 m}{27}\(O_3\langle C_1\rangle + C_2\langle O_1\rangle\)
-\frac{m^2}{54} O_3\langle O_1\rangle, \\
\label{Zeff-m2}
S_{\chi{\rm-odd}}(m)&=&\frac{\lambda_2 m}{27}\( C_1\langle O_3\rangle+ O_1\langle C_2\rangle\)
-\frac{m^2}{54} O_1\langle O_3\rangle.
\eea
Let us note that the terms linear in the mass parameter $m$ are also linear in the chirality-odd
operators $C_{1,2}$.

With this addition to the previous effective action \ur{Seff-2D} we now turn to solving
the self-consistency equations for the operator averages $\langle O_{1,2,3}\rangle$
and $\langle C_{1,2}\rangle$, $\langle \overline C_{1,2}\rangle$.
At $m\neq 0$ there is a solution for the chiral condensates $\langle C_{1,2}\rangle$ in the whole
$(\lambda_1,\lambda_2)$ plane (we still keep for simplicity $\lambda_3=0$). However, the dependence
of the chiral condensates on the mass parameter $m$ is totally different depending on whether
we are in the region I where chiral symmetry is broken, or in the region II where it is preserved.

In the region II the dependence of the chiral condensates on the mass is linear at small $m$;
if $m$ goes to zero the chiral condensates vanish. In the region I the dependence of the chiral
condensates on the small parameter $m$ that breaks the symmetry explicitly is {\em non-analytic}.
Actually the chiral condensates are proportional to the sign functions of $m$,
$\langle C_{1,2}\rangle \sim {\rm sign}(m)$, see Fig.~5. The behavior of the chiral condensate
$\langle C_{1,2}\rangle$ in the whole $(\lambda_1,\lambda_2)$ plane (at $\lambda_3=0$) is shown in Fig.~6.

\begin{figure}[t]
\begin{center}
\begin{minipage}[h]{0.45\textwidth}
\includegraphics[width=0.9\textwidth]{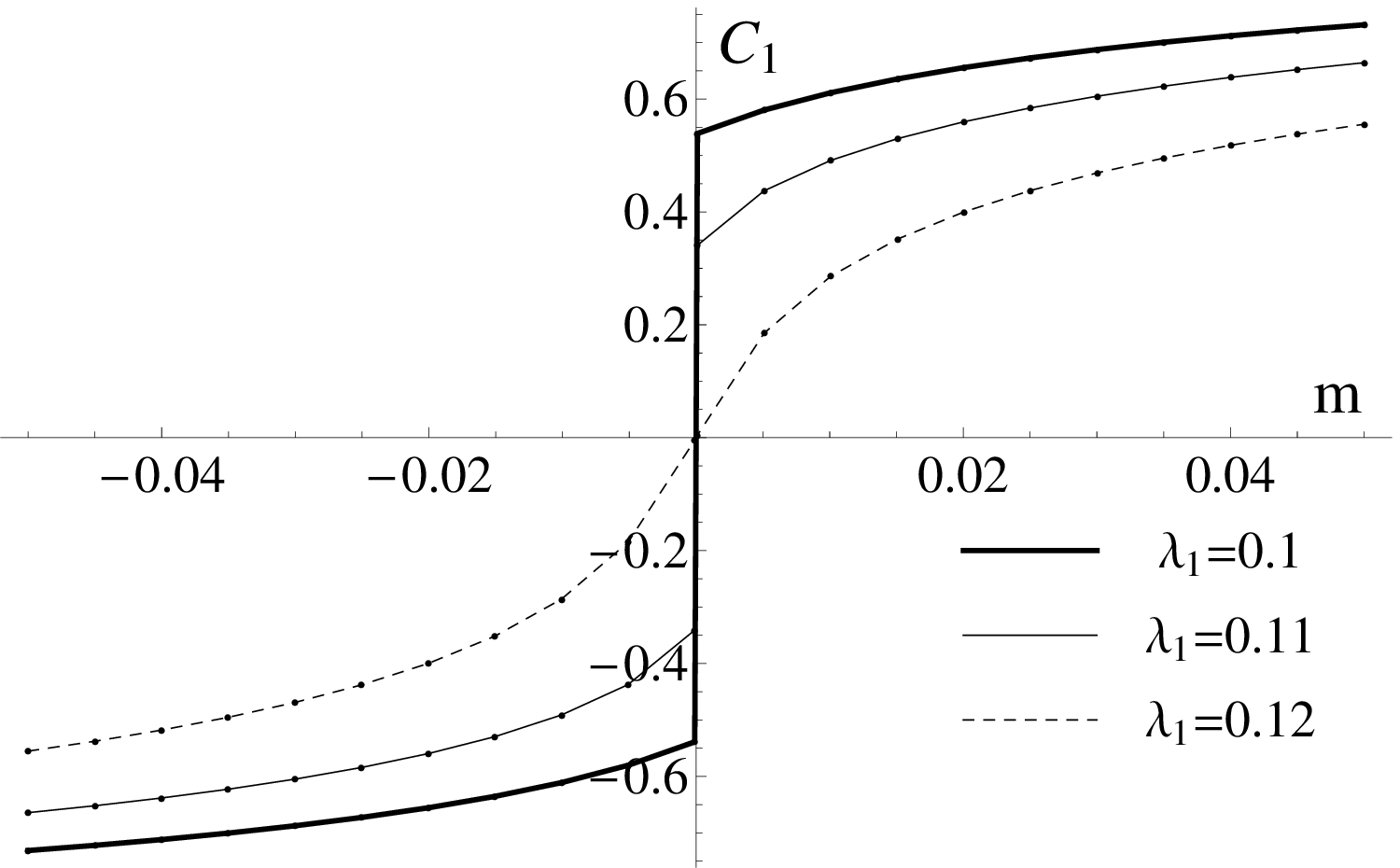}
\caption{The dependence of the chiral condensate $\langle C_1\rangle $ on the mass parameter $m$
with the varying value of $\lambda_1$ at fixed value of $\lambda_2$. The step-like behavior
$\sim{\rm sign}(m)$ signals the spontaneous breaking of symmetry. In this example, $\lambda_1=0.12$
is the 2$^{\rm nd}$ order phase transition point where the chiral condensate abruptly vanishes when $m=0$.}
\end{minipage}
\hfill
\begin{minipage}[h]{0.45\textwidth}
\includegraphics[width=1\textwidth,bb= 30 10 400 340, clip=true]{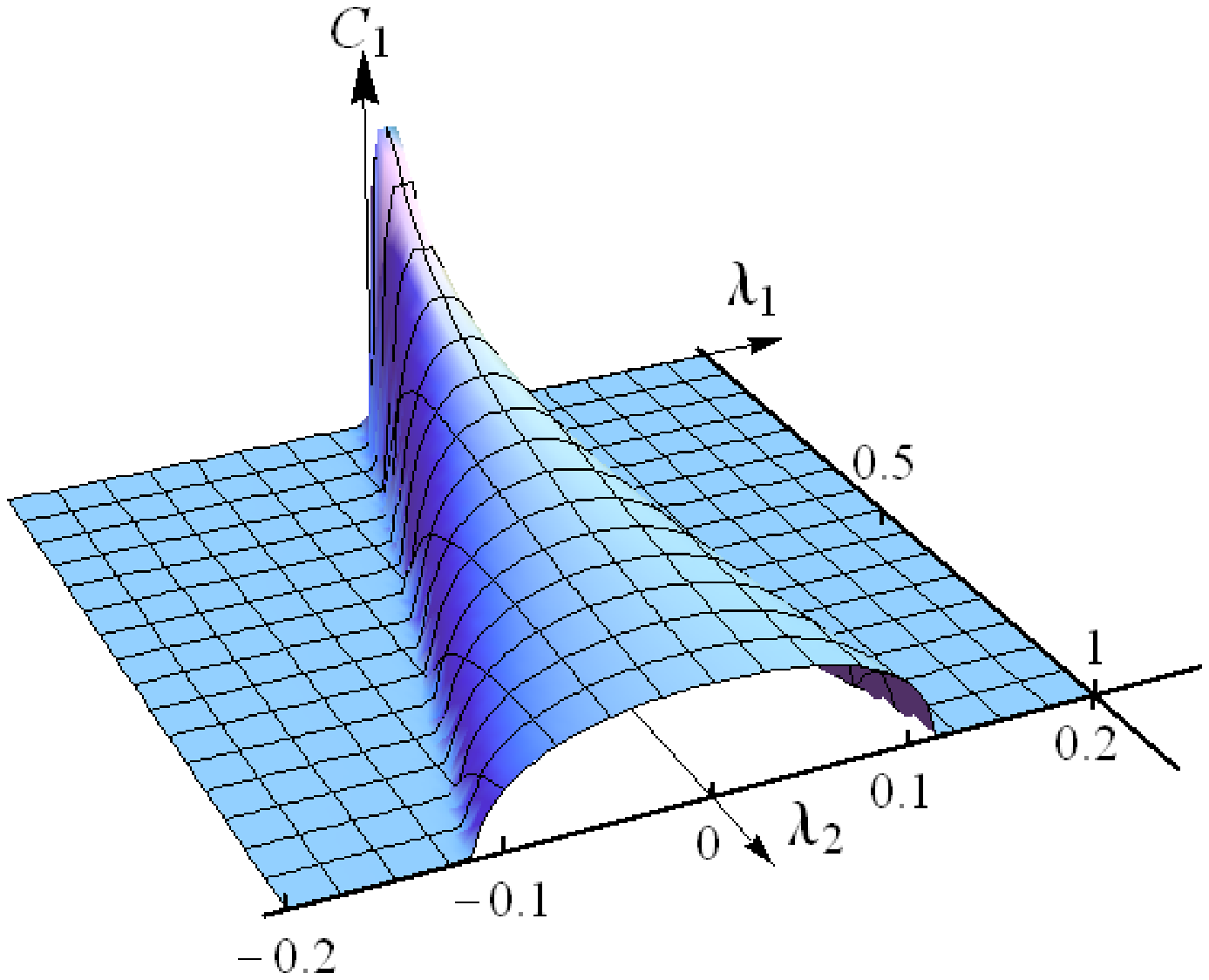}
\caption{The value of the chiral condensate $\langle C_1\rangle $ in the $(\lambda_1,\lambda_2)$ plane.
At the phase transition line $|\lambda_2|=8.69\,|\lambda_1|$ the condensate vanishes with
an infinite first derivative.}
\end{minipage}
\end{center}
\end{figure}

Figs.~5,6 clearly demonstrate that there is a range of coupling constants where the theory undergoes
spontaneous breaking of the continuous $U(1)_A$ chiral symmetry, with a line of the 2$^{\rm nd}$ order
phase transition separating the phases.

\subsection{No fermion number violation}

The effective action (\ref{Seff-2D}) contain operators $W_{1,2}$ violating the $U(1)_V$ symmetry
\ur{vector} related to the fermion number conservation. If $\langle W_{1,2}\rangle \neq 0$ it signals
the spontaneous violation of this symmetry. Fermion number conservation is spontaneously broken {\it e.g.}
in ordinary superconductors and ``color'' superconductors in QCD. However, in contrast to chiral symmetry
that is also broken in QCD, spontaneous fermion condensation usually happens not in the vacuum but
at nonzero chemical potential for fermions, since interactions are effectively amplified near the
Fermi surface. In this Subsection we look for the spontaneous fermion number non-conservation in the same
$2d$ model where we observe spontaneous breaking of the $U(1)_A$ symmetry in the mean-field approximation.

Following the same logic as in the previous Subsection we introduce a small action term that violates
the $U(1)_V$ symmetry explicitly,
\beq\nn
S_{\rm B-odd}=\int d^2x \det(\tilde e)\, b\, \psi^1\psi^2.
\eeq
This operator preserves the chiral $U(1)_A$ symmetry. The correction to the effective one-triangle action is:
\bea\nn
S_{\rm B-odd}(m,n)&=&\frac{\lambda_2 b}{27}\( W_2\langle O_1\rangle+ O_3\langle W_1\rangle\), \\
\nn
S_{\rm B-odd}(m)&=&\frac{\lambda_2 b}{27}\( O_1\langle W_2\rangle + W_1\langle O_3\rangle\).
\eea
We solve again the self-consistency equations on the operator averages but now with this addition,
and look for non-analytic dependence on the small parameter $b$. In contrast to the case of
spontaneous chiral symmetry breaking we do not find such solutions in the whole parameter
space $\Lambda=(\lambda_1,\lambda_2,\lambda_3)$.

We conclude that the fermion number conservation is not broken spontaneously in the model, except
maybe along the line of the chiral phase transition. We did not study the inclusion of a chemical
potential for fundamental fermions -- that would explicitly violate Lorentz symmetry but presumably
make the phase diagram of the model more rich.

There are no reasons why fermion number conservation would not break spontaneously, say, in $4d$,
and the mean field method suggested here is a simple way to detect it.

\subsection{Full phase diagram}

The full action compatible with the principles proclaimed has in $2d$ three terms and
consequently three coupling constants. In the previous Subsections we have restricted
our study to the case of $\lambda_3=0$.

Actually we repeat all the steps described above also for $\lambda_3\neq 0$. The algebra becomes
more cumbersome but still doable. We find that the chiral symmetry breaking phase I occupies the cone
\beq
\lambda_2^2 < 77.23\, \lambda^2_1 +5.36\, \lambda_3^2
\la{cone}\eeq
shown in Fig.~7; Fig.~4 is its section at $\lambda_3=0$. We remark that the accuracy of the mean-field
approximation for some reason deteriorates as $\lambda_3$ grows. Still the exact relation \ur{<S>}
holds even at $\lambda_3\to \infty$ up to a factor of 1.6.
\begin{figure}[h]
\begin{center}
\includegraphics[width=0.4\textwidth,bb= 125 100 600 600, clip=true]{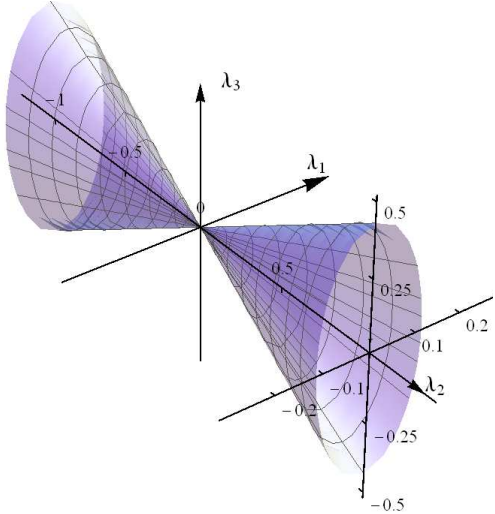}
\caption{Spontaneous chiral symmetry breaking takes place inside a cone in the full parameter
space $\Lambda=(\lambda_1,\lambda_2,\lambda_3)$. }
\end{center}
\end{figure}

\section{Low-energy action for propagating fields}

The theory defined by the partition function \ur{Z} is in fact ultra-local: all correlation functions of
gauge-invariant operators generally decay exponentially at the separation of a few lattice cells.
This is clear on general grounds but we have also checked it by numerical simulations on a $2d$
lattice of limited volumes. Special measures should be taken to ensure that certain degrees
of freedom propagate to distances that are large in lattice units. The situation here is
different from the common lattice gauge theory where it is sufficient to take the limit $\beta\to \infty$
where $\beta$ is the inverse gauge coupling to guarantee long-range correlations in lattice units.
In our theory there is no such obvious handle.

However there are ways to {\em guarantee} that long-range correlations appear; moreover that can
be checked in the mean-field approximation. An example which we consider here is provided by the
Goldstone theorem: If a global continuous symmetry is broken spontaneously the associated Goldstone
bosons are exactly massless and hence propagate to large distances.

In the previous Section we have shown that the continuous $U(1)_A$ or chiral symmetry is spontaneously
broken in a broad range of the space of the coupling constants. Supposing the coupling constants
are chosen inside that range (inside the cone in Fig.~7), there is a massless Goldstone excitation
$\alpha(x)$ being the phase of the $U(1)_A$ rotation \ur{axial}.

Under this rotation the chirality-violating operators $C_{1,2}$ and $\overline C_{1,2}$ transform
as
\beq
C_1^\pm=C_1\pm \overline C_1\rightarrow e^{\pm i\alpha}C_1^\pm,
\qquad\qquad
C_2^\pm=C_2\pm \overline C_2\rightarrow e^{\pm i\alpha}C_2^\pm.
\la{C-transf}\eeq
To derive the low-energy action for the Goldstone field we allow the phase $\alpha$ to vary slowly from cell to cell:
\beq
\langle C_{1,2}^{\pm}\rangle = \rho_{1,2}e^{\pm i\alpha(\text{cell})}.
\la{C-param}\eeq
We parameterize the operator averages $\langle C_{1,2}^{\pm}\rangle$ in the same way and re-derive
the effective action \ur{Seff-2D} for the fields inside the triangle cavity, taking now into account
that the operator averages have slightly different phases in the cells surrounding the cavity. Then,
integrating over the fields inside the cavity we find the effective one-cell partition function ${\cal Z}_1$
modified by the varying nearest neighborhood. If $\alpha$ is the same for all neighboring cells
it is the same expression as in Section V, let us call it ${\cal Z}_{10}$. However, there will be
further terms depending on the gradients of $\alpha(x)$, we are now after.

The full partition function is, in the mean field approximation, a product of ${\cal Z}_1$'s over all cells
whose number is $M$. Therefore the action for the Goldstone field $\alpha(x)$ is
\bea\la{SG1}
S_G &=& -M\ln{\cal Z}_1 = -M\ln {\cal Z}_{10}\\
\n
&-& M\left[\frac{1}{{\cal Z}_{10}}\frac{\partial{\cal Z}_1}{\partial \alpha_i}\Delta \alpha_i
+\frac{1}{2{\cal Z}_{10}}\left(\frac{\partial^2{\cal Z}_1}{\partial\alpha_i\partial \alpha_j}
-\frac{1}{{\cal Z}_{10}}\frac{\partial{\cal Z}_1}{\partial \alpha_i}\frac{\partial{\cal Z}_1}{\partial \alpha_j}\right)
\Delta \alpha_i\Delta \alpha_j
+{\cal O}(\Delta\alpha^3)\right]
\eea
where $\alpha_i$ is the value of the phase attributed to one of the 12 neighbor cells $i$, and
$\Delta\alpha_i$ is the difference between $\alpha_i$ and $\alpha_0$ attributed to the central
cavity cell; the summation goes over all neighbor cells. It is important that the dependence
of ${\cal Z}_1$ on $\alpha_i$ starts from quadratic terms, which is the consequence of chiral symmetry;
hence $\partial{\cal Z}_1/\partial\alpha_i=0$, and we are left with second derivatives.

Ignoring the first $\alpha$-independent term in \Eq{SG1} we find that the action is quadratic
in the jumps $\Delta\alpha$ from one cell to the neighbor ones,
\beq
S_G=-M\frac{1}{2{\cal Z}_{10}}\frac{\partial^2{\cal Z}_1}{\partial\alpha_i\partial \alpha_j}
\Delta \alpha_i\Delta \alpha_j
+ {\cal O}(\Delta\alpha^3).
\la{SG2}\eeq

We now introduce a coordinate system by mapping the centers of the cells to coordinates $x^\mu(i)$
(Section IV). If the changes of $\alpha$ from a cell to neighbor cells are small we can expand
\beq
\Delta\alpha_i = \partial_\mu \alpha \Delta x^\mu_i
+\frac{1}{2}\,\partial_\mu\partial_\nu \alpha \Delta x^\mu_i\Delta x^\nu_i+\ldots
\la{alpha-Taylor}\eeq
where $\Delta x^\mu_i = x^\mu(i) - x^\mu(0)$ is the distance between the coordinate attributed
to the cell $i$ and that attributed to the cavity cell, in a given coordinate frame $x^\mu(i)$.
Putting this expansion into \Eq{SG2} we obtain
\beq
S_G=-M\,\frac{1}{2{\cal Z}_{10}}\,\frac{\partial^2{\cal Z}_1}{\partial\alpha_i\partial \alpha_j}\,
\Delta x^\mu_i\Delta x^\nu_j\; \partial_\mu \alpha \partial_\nu \alpha + {\cal O}(\Delta x^3).
\la{SG3}\eeq
The first factor $M$, the full number of cells on the lattice can be written as
\beq
M=\sum_{\rm cells} = \int \frac{d^2x}{V(\text{cell})}
\la{summ}\eeq
where $V(\text{cell})$ is the cell volume in a given frame, see \Eq{V_simplex-1}. The combination
\beq
\lim_{\Delta x\to 0}\;\frac{1}{V(\text{cell})}\,\frac{1}{{\cal Z}_{10}}\,
\frac{\partial^2{\cal Z}_1}{\partial\alpha_i\partial \alpha_j}\,\Delta x^\mu_i\Delta x^\nu_j
\;\bydef\;-\sqrt{G}\, G^{\mu\nu}
\la{G}\eeq
transforms under the change of the map $x^\mu\to x'^\mu(x)$ as a product of the contravariant
tensor times the square root of the determinant of a covariant tensor, hence the notations
in the right hand side or \Eq{G}. Its particular form depends, of course, on the coordinate
system chosen. For a concrete map to the Cartesian coordinates of the lattice drawn
in Fig.~3 we find that it is proportional to a unity tensor,
\beq
\sqrt{G}G^{\mu\nu}\Big|_{\text{regular lattice}}=T(\lambda_{1,2,3}) \delta^{\mu\nu}
\la{T}\eeq
where the proportionality coefficient $T(\lambda_{1,2,3})$ is shown in Fig.~8; it is proportional
to a combination of the moduli of the chiral condensates $\rho_{1,2}$, see \Eq{C-param}. This result is
in conformity with the average flatness of the space found in Section VI.A. If one chooses
another coordinate map $\sqrt{G}G^{\mu\nu}$ changes accordingly.
\begin{figure}[h]
\begin{center}
\includegraphics[width=0.4\textwidth]{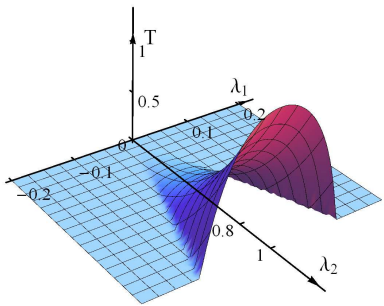}
\caption{The normalization factor $T(\lambda_1,\lambda_2,0)$ in the low-energy effective chiral Lagrangian, \Eq{T}.}
\end{center}
\end{figure}

We thus arrive at a diffeomorphism-invariant low-energy action for the massless Goldstone field:
\beq
S_G = \frac{1}{2}\int d^2x\,\sqrt{G}G^{\mu\nu}\,\partial_\mu \alpha \partial_\nu \alpha\,.
\la{SG4}\eeq
This field can propagate infinitely far in lattice units since its masslessness is guaranteed
by the Goldstone theorem.

To complete our study of the spontaneous chiral symmetry breaking we derive the analog
of the Gell-Mann--Oakes--Renner relation for the pion mass in QCD, expressed through the
quark masses. If chiral symmetry is broken explicitly by a small fermion mass term \ur{mass-term}
the phase of the chiral condensate becomes a pseudo-Goldstone field with a mass proportional
to the {\em square root} of the fermion mass $m$.

Indeed, the addition of the mass term \ur{mass-term} changes the effective one-cavity
partition function:
\beq
{\cal Z}_1\to {\cal Z}_1+m {\cal Z}_m+\mathcal{O}(m^2).
\la{Z-mass}\eeq
The ${\cal Z}_m$ piece depends explicitly on the chiral condensate phase $\alpha$ introduced
in \Eq{C-param}, and from symmetry considerations it is clear that the expansion starts
from the $\alpha^2$ term; direct calculation confirms it.

Summing up the mean-field action over the whole lattice one uses the relation \ur{summ} where
the r.h.s. behaves as $\sim\!\!\int d^2x\,\sqrt{G}$ according to the transformation properties
under the change of the coordinate system $x^\mu(i)$ attributed to the lattice. We obtain thus
the action for the pseudo-Goldstone mode in the continuum limit
\beq
S_G = \frac{1}{2}\int d^2x\,\sqrt{G}\left(G^{\mu\nu}\,\partial_\mu \alpha \partial_\nu \alpha
+\mu^2\,\alpha^2\right)+{\cal O}(\partial^2\alpha\partial^2\alpha)+{\cal O}(\alpha^4),
\qquad \mu^2\sim m,
\la{SG5}\eeq
where $\mu$ is proportional to the pseudo-Goldstone boson mass. We see that it is proportional
to the {\em square root} of the mass parameter $m$ that breaks chiral symmetry explicitly.
In QCD, this is known as the Gell-Mann--Oakes--Renner relation for the pion mass. The coefficient
in this relation depends on the coupling constants $\lambda_{1,2,3}$. At the 2$^{\rm nd}$ order
phase transition surface of the cone in Fig.~7 the pseudo-Goldstone mass goes to zero at fixed $m$.
\vskip 0.3true cm

There is a famous Mermin--Wagner theorem stating that a continuous symmetry cannot be spontaneously
broken in $2d$ as the resulting Goldstone bosons would have an unacceptably large, actually
divergent free energy. Since the mean field approximation misses the Goldstone physics, one can
argue that the spontaneous chiral symmetry breaking we observe is an artifact of the approximation.
If, however, the Goldstone field $\alpha(x)$ is Abelian as here, the actual phase is, most likely,
that of Berezinsky--Kosterlitz--Thouless where the chiral condensate $\rho\,e^{i\alpha}$ indeed
vanishes owing to the violent fluctuations of $\alpha(x)$ defined on a circle $(0,2\pi)$,
but the correlation functions of the type $\langle e^{i\alpha(x)}\,e^{-i\alpha(y)}\rangle$
have a power-like behavior, and there is a phase transition depending on the original couplings
of the theory.

In any case our primary goal here is to learn how to deal with the lattice regularized spinor quantum
gravity which is a new type of a theory. The mean-field approximation is one possible approach that
is expected to work even better in higher dimensions where, as a matter of fact, the Mermin--Wagner
theorem does not apply.

\section{How to obtain Einstein's limit?}

The apparent diffeomorphism-invariance of \Eq{SG5} is built in by our construction of the
lattice and lattice action in Section IV. As soon as there are degrees of freedom that
can propagate to long distances, their low-energy effective action is diffeomorphism-invariant
in the continuum limit.

In the previous Section the appearance of long-propagating mode has been guaranteed by the
Goldstone theorem. However, it concerns only the specific Goldstone modes associated with
the spontaneous breaking of continuous symmetry. Other degrees of freedom remain heavy: their
correlation functions decay exponentially after a few lattice cells. If one attempts to write
an effective low-energy action for the classical metric tensor $g^{\rm cl}_{\mu\nu}$
(see below its exact definition) it will have the diffeomorphism-invariant form,
\beq
S_{\rm low}=\int dx \,\sqrt{g^{\rm cl}}(-c_1+c_2 R(g^{\rm cl})+\ldots),
\la{Slow}\eeq
with the constants $c_{1,2}$ computable, in principle, from the original coupling constants of
the lattice-regularized theory. However, if one does not take special measures, the ratio
$\sqrt{c_1/c_2}$, playing the r\^ole of the graviton mass, will be on the order of the inverse
lattice spacing. In such a situation it is senseless to introduce the metrics in the first place.
It makes sense only if $\sqrt{c_1/c_2}$ happens to be zero or very small, such that the graviton
and the Newton force propagates to large distances.

To ensure it, it is sufficient to stay,{\it e.g.}, at the phase transition of the second order,
where all degrees of freedom become massless. The classical metric tensor $g^{\rm cl}_{\mu\nu}$
and the effective action functional $\Gamma[g^{\rm cl}_{\mu\nu}]$ can be introduced by means of the
Legendre transform (proposed in this context also by Wetterich~\cite{Wetterich:2011xv}). One introduces
first the generating functional for the stress-energy tensor $\Theta^{\mu\nu}$ as an external source,
\beq
e^{W[\Theta]}=\int d\psi^\dagger\,d\psi\,d\omega_\mu\,\exp\left(S+\int \hat g_{\mu\nu}\,\Theta^{\mu\nu}\right),
\la{GF}\eeq
where $S$ the fermionic action and $\hat g_{\mu\nu}$ is a 4-fermion operator built from the frame
fields \urs{e-def}{f-def} or, after discretization, from their lattice versions \urs{tilde-e}{tilde-f}.
The classic metric  field is by definition
\beq
g^{\rm cl}_{\mu\nu}\bydef\langle\hat g_{\mu\nu}\rangle =\frac{\delta W[\Theta]}{\delta \Theta^{\mu\nu}}.
\la{gclass}\eeq
This equation can be solved to give the functional $\Theta^{\mu\nu}[g^{\rm cl}]$.
Using it one can construct the effective action as the Legendre transform:
\beq
\Gamma[g^{\rm cl}]=W[\Theta]-g^{\rm cl}_{\mu\nu}\,\Theta^{\mu\nu}.
\la{Gamma}\eeq
At the phase transition fluctuations are long-ranged. For long-range fluctuations it is legal
to take the continuum limit of the lattice, which is diffeomorphism-invariant. The low-energy
limit of diffeomorphism-invariant actions for a quantity transforming as a metric tensor is
uniquely given by \Eq{Slow}. Moreover, the cosmological term necessarily has zero coefficient,
$c_1=0$, since otherwise the graviton would propagate to a finite distance $\sqrt{c_2/c_1}$,
which contradicts the masslessness of the fluctuations at the phase transition. This is how
one can recover Einstein's gravity from the lattice-regularized spinor theory.

In principle, the effective Einstein--Hilbert action from spinor quantum gravity can be derived
in the mean-field approximation similarly to our derivation of the low-energy effective chiral
Lagrangian in Section VII. However, in $2d$ where we have so far succeeded in developing the
mean-field method the Einstein--Hilbert action is a full derivative and there are no gravitons
or the Newton force. Therefore, the derivation of \Eq{Slow} has to be postponed till higher
dimensions are studied along the lines of the present paper.

\section{Dimensions}

In this paper, we use unconventional dimensions of the fields, which, however, we believe are
natural and adequate for a microscopic theory of quantum gravity. The fermion fields are normalized
by the Berezin integral \ur{Berezin} and are dimensionless, hence the composite frame field
\ur{e-def} has the dimension 1/length and the metric tensor has the dimension 1/length$^2$,
in contrast to the conventional dimensionless metric tensor. On the other hand all diffeomorphism-invariant
quantities are dimensionless in our approach. In this section we explain why it is convenient,
and what is the relation to the usual approach.

The historic tradition in General Relativity is that the space-time at infinity is flat, therefore
one can safely choose the coordinate system such that $g_{\mu\nu}$ is a unity matrix there.
This sets the traditional dimensions of the fields. In particular, the scalar curvature has
the dimension 1/length$^2$, the fermion fields have the dimension 1/length$^{\frac{3}{2}}$, {\it etc.}
However, in a diffeomorphism-invariant quantum theory where one can perform arbitrary change
of coordinates $x^\mu \to x'^\mu(x)$ not necessarily identical at infinity, for example, a dilatation
$x^\mu \to x^\mu/\rho$, and where $g_{\mu\nu}$ can {\it a priori} strongly fluctuate at infinity,
this convention is not convenient.

The natural dimensions of the fields are those that are in accordance with their transformation
properties: any contravariant vector transforms as $x^\mu$ and has the dimension of length,
a covariant vector, in particular the frame field $e_\mu$ transforms as a derivative and has
the dimension 1/length, $g_{\mu\nu}$ has the dimension 1/length$^2$, {\it etc.}
World scalars like the scalar curvature and the fermion fields are, naturally, dimensionless.
In fact it is a tautology: a quantity invariant under diffeomorphisms is in particular
invariant under dilatations and hence has to be dimensionless.

In this convention, any diffeomorphism-invariant action term is by construction dimensionless and
is accompanied by a dimensionless coupling constant, as in \Eq{Z}.

Let us suppose that we have a microscopic quantum gravity theory at hand that successfully generates
the first terms in the derivative expansion of the effective action,
\beq
\Gamma = -c_1\int d^4x\,\sqrt{g}+c_2\int d^4x\,\sqrt{g}R+\ldots
\la{1}\eeq
where $c_{1,2}$ are certain dimensionless constants expressed through the dimensionless couplings
$\lambda_{1,2,\ldots}$ of the original microscopic theory. The ground state of the action (1) is
the space with constant curvature $R=2c_1/c_2$, represented {\it e.g.} by a conformal-flat metric
\beq
g_{\mu\nu}=\frac{6c_2}{c_1}\,\left(\frac{2\rho}{\left((x-x_0)^2+\rho^2\right)}\right)^2\,\delta_{\mu\nu}
\la{2}\eeq
where $x_0$ and $\rho$ are arbitrary. At the vicinity of some observation point $x_0$ it can be made
a unity matrix by rescaling the metric tensor,
\beq
g_{\mu\nu}=m^2\,\bar g_{\mu\nu},\qquad \bar g_{\mu\nu} =\delta_{\mu\nu},\qquad
m = \sqrt{\frac{6c_2}{c_1}}\,\frac{2}{\rho}
\la{3}\eeq
where the rescaling factor $m$ has the dimension of mass, and $\bar g_{\mu\nu}$ has the conventional
zero dimension. At this point one can rescale other fields to conventional dimensions, in particular
introduce the new fermion field $\bar\psi$ of conventional dimension $m^{3/2}$:
\beq
\psi = m^{-3/2}\,\bar\psi,\qquad \psi^\dagger = m^{-3/2}\,\bar\psi^\dagger\,.
\la{4}\eeq
The new composite dimensionless tetrad field compatible both with  Eqs. \ur{3} and \ur{4} is
\beq
\bar e^A_\mu = \frac{1}{m}\,e^A_\mu = \frac{1}{m^4}\,i\left(\bar\psi^\dagger\gamma^A\nabla_\mu \bar\psi
+\bar\psi^\dagger\overleftarrow\nabla_\mu\gamma^A\bar\psi\right).
\la{5}\eeq
One can now rewrite the action \ur{1} together with the fermionic matter in terms of the new rescaled fields
denoted by a bar,
\beq
S=-\underbrace{c_1\,m^4}_{2\Lambda =\lambda^4}\int d^4x\,\sqrt{\bar g}
+\underbrace{c_2\,m^2}_{M_{\rm P}^2=1/\sqrt{16\pi G_{\rm N}}}\int d^4x\,\sqrt{\bar g}\bar R
+m^0\int d^4x\,\sqrt{\bar g}\,\bar e^{A\mu}\left(\bar\psi^\dagger\gamma^A\nabla_\mu\bar\psi +{\rm h.c.}\right).
\la{6}\eeq
Underbraced are the cosmological constant and the Plank mass squared, respectively; numerically,
$\lambda=2.39\cdot 10^{-3}\;{\rm eV}$, $M_{\rm P}=1.72\cdot 10^{18}\;{\rm GeV}$. The dimensionless
ratio of these values,
\beq
\frac{\lambda}{M_{\rm P}}=\left(\frac{c_1\,m^4}{c_2^2\,m^4}\right)^{\frac{1}{4}}
=\left(\frac{c_1}{c_2^2}\right)^{\frac{1}{4}}=1.39\cdot 10^{-30}\,,
\la{7}\eeq
is the only meaningful quantity in pure gravity theory, independent of the arbitrary scale
parameter $m$. If a fermion obtains an effective mass, {\it e.g.} as a result of the spontaneous chiral
symmetry breaking, leading to an additional term in the effective low-energy action
\beq
S_m=\int d^4x\,\sqrt{g}\,\psi^\dagger\,{\cal M}\,\psi
= \underbrace{m\,{\cal M}}_{{\rm fermion\;mass}\;m_f}\int d^4x\,\sqrt{\bar g}\,\bar\psi^\dagger\bar\psi\,,
\la{8}\eeq
then the ``theory of everything'' has to predict also other dimensionless ratios. For example, taking
the top quark mass $m_t=172\;{\rm GeV}$ one has to be able to explain the ratio
\beq
\frac{m_t}{\sqrt{\lambda M_{\rm P}}}=\frac{{\cal M}}{c_1^{\frac{1}{4}}c_2^{\frac{1}{2}}}=0.0848\,.
\la{9}\eeq

In other words, one can measure the Newton constant (or the Planck mass) or the cosmological constant
in units of the quark or lepton masses or the Bohr radius. Only dimensionless ratios make sense and
can be, as a matter of principle, calculated from a microscopic theory. To that end it is convenient
and legitimate to use natural dimensions when $g_{\mu\nu}$ has the dimension 1/length$^2$ whereas all world
scalars are dimensionless, be it the scalar curvature $R$, the interval $ds$, the fermion field $\psi$
or any diffeomorphism-invariant action term.

\section{Conclusions}

We have formulated a lattice-regularized spinor quantum gravity that is well defined and well
behaved both for large-amplitude and high-frequency fluctuations. In any number of dimensions
one can construct a variety of fermionic actions that are invariant ({\it i}) under local
Lorentz transformations and ({\it ii}) under diffeomorphisms in the continuum limit.
We have built quite a few action terms satisfying ({\it i}) and ({\it ii}) for any number
of dimensions. In fact our list of possible fermionic action terms can be expanded further
if some of the additional requirements are relaxed. Therefore, we actually formulate a whole
class of new kind of theories in any number of dimensions, characterized by a set of dimensionless
coupling constants $\lambda_{1,2,\ldots}$.

The continuum limit shows up if all degrees of freedom or at least some of them are slowly
varying fields from one lattice cell to another. This is, generally, not fulfilled: generically,
all correlation functions decay exponentially over a few lattice cells. For such ``massive''
degrees of freedom the theory is at the ``strong coupling'' regime where the continuum limit
is not achieved and remains dormant.

There must be special physical reasons for massless excitations in the theory, for which the
continuum limit makes sense and diffeomorphism-invariance becomes manifest. One such reason
is spontaneous breaking of continuous symmetry where the existence of massless fields
is guaranteed by the Goldstone theorem. To show that spontaneous breaking may be typical in
such kind of theories, we have developed a new mean-field approximation. We have checked its
accuracy in a $1d$ exactly solvable model, and in a full $2d$ theory where certain exact relations
can be derived. The exact relations tell us nice things: the physical or invariant volume
occupied by the system is extensive as due to the non-compressibility of fermions, the volume
variance (or susceptibility) is also extensive showing that it is a true quantum vacuum, and the
average curvature is, at least in $2d$, zero meaning that the quantum space is on the average flat.
They also tell us that our mean-field approximation is rather accurate, and the accuracy can be
systematically improved.

We show, within the mean-field method, that the spontaneous breaking of chiral symmetry happens
in a broad range of the coupling constants and that in this range the low-energy action for
the Goldstone field (or pseudo-Goldstone if we add a term explicitly breaking symmetry) is
diffeomorphism-invariant, as expected.

To obtain the low-energy Einstein limit one has to stay at the second-order phase transition
surface in the space of the coupling constants. There the masslessness of excitations, and not
only of the Goldstone ones, is guaranteed. Hence one can go to the continuum limit where the
diffeomorphism-invariance is also guaranteed by construction. Therefore, we expect that the
effective low-energy action for the classical metric tensor, derived through the Legendre transform,
is just the Einstein--Hilbert action, with the zero cosmological term. This can be probably seen
already in the mean-field approach for dimensions higher than two. This work is in progress.

\vskip 0.7true cm

\noindent
{\bf Acknowledgments.}

We thank Victor Petrov, Maxim Polyakov and Alexander Tumanov for helpful discussions.
This work has been supported in part by Russian Government grant NS4801.2012.2, by Deutsche
Forschungsgemeinschaft (DFG) grant 436 RUS 113/998/01, and by the BMBF grant 06BO9012.

\appendix

\section{Mean field method in a $1d$ model}

We consider here a $1d$ toy model with fermions and $U(1)$ gauge symmetry, that
can be solved exactly. We then apply the mean field method to this model to check how accurately
does it reproduce the exact solution. We obtain quite satisfactory results.

We take the same fields as in the full $2d$ model, namely the doublet of fermion fields
$\psi^\alpha,\;\psi^\dagger_\alpha,\;\alpha=1,2$, that transform under the $U(1)$ gauge transformation
as $\psi\to V\psi,\;\psi^\dagger \to \psi^\dagger V^\dagger,\;V=\exp(i\alpha\sigma_3)$. The gauge
field is represented by link variables $U_{ij}=\exp(-i\omega_{ij}\sigma^3/4)$ that transform as
$U_{ij}\to V_i U_{ij}V^\dagger_j$.

We construct the lattice version of the two ``frame'' fields, as in \Eqs{tilde-e}
{tilde-f},
\bea\n
\tilde e_{i,j}^A &=& i(\psi^\dagger_j U_{ji}\sigma^A U_{ij}\psi_j-\psi^\dagger_i \sigma^A\psi_i),\quad A=1,2,\\
\tilde f_{i,j}^A &=& \psi_i^\dagger \sigma^A U_{ij}\psi_j-\psi^\dagger_jU_{ji}\sigma^A\psi_i,
\la{eh-lattice}\eea
(which however do not have the meaning of frame fields in $1d$), and form the action that is quite
similar to the full $2d$ action \ur{tildeS0-number2}:
\beq
S=\sum_{i=1}^N\left[\frac{\lambda_1}{8}(\tilde e^A_{i,i+1}\tilde e^A_{i,i+1}
+\tilde e^A_{i+1,i}\tilde e^A_{i+1,i})
+\frac{\lambda_2}{4}\tilde f^A_{i,i+1}\tilde f^A_{i,i+1}
+2\mu (\psi^\dagger_i\psi_i)^2(\psi^\dagger_{i+1}\psi_{i+1})^2\right]
\la{S-1d}\eeq
where $\lambda_{1,2}$ and $\mu$ are the coupling constants. The partition function is defined as
a product of Berezin integrals on the $1d$ lattice with $N$ points:
\beq
{\cal Z}=\prod_{i=1}^N\int\!d\psi^1_id\psi^2_i d\psi^\dagger_{i\,1}d\psi^\dagger_{i\,2}dU_{i,i+1}\,e^S.
\la{Z-1d}\eeq
We imply antiperiodic boundary conditions for fermion fields and periodic boundary conditions for link
variables.

The partition function \ur{Z-1d} is {\em exactly computable} by a kind of transfer-matrix method.
Diagonalizing the transfer matrix we obtain a nontrivial result:
\beq
{\cal Z}=2(1+(-1)^N)\lambda_2^N+\(\lambda_1-\sqrt{2}\sqrt{\lambda_1^2+\lambda_2^2+\mu}\)^N
+\(\lambda_1+\sqrt{2}\sqrt{\lambda_1^2+\lambda_2^2+\mu}\)^N.
\la{Z-1d-1}\eeq
The fact that the partition function has the form of a sum of extensive exponents means that actually
it describes simultaneously four independent phases or states of the system that do not compete and hence
do not mix up in the thermodynamic limit $N\to \infty$. The stable phase is the one with the lowest free energy,
that is with the largest partition function.

Depending on the relation between the coupling constants one of the terms in \Eq{Z-1d-1} prevails at $N\to\infty$:
\begin{itemize}
\item{Phase 0:}  $|\lambda_2| > |\lambda_1|$ and $-(\lambda_1^2+\lambda_2^2)
< \mu <-\frac{1}{2}\(\lambda_1\pm\lambda_2\)^2$; the partition function
is given by the first term
\item{Phase 1:}  $\mu >-\frac{1}{2}\(\lambda_1+\lambda_2\)^2$,
$\lambda_1 < 0$; the partition function is given by the second term
\item{Phase 2:}  $\mu >-\frac{1}{2}\(\lambda_1+\lambda_2\)^2$,
$\lambda_1 > 0$; the partition function is given by the third term
\item{Phase 3:}  $\mu <-\(\lambda_1^2+\lambda^2_2\)$, $|\lambda_1| > |\lambda_2|$;
the partition function is complex and has no smooth thermodynamic limit.
\end{itemize}
In phase 0 fermions in the neighbor lattice sites form an ordered state of the type ``pair-gap-pair-gap...''
all over the lattice, where the ``pair'' means that there are four link matrices in the integration over link variables,
and ``gap'' means zero matrices. It can be realized only on even-$N$ lattices, hence it is a lattice artifact,
and we do not consider it further. Phases 1 and 2 are states where two link matrices appear in all link integrations.
We concentrate of phases 1 and 2 only in what follows.

We also calculate exactly average values of the following 4-fermion operators:
\bea\la{ee-1}
\langle\tilde e_{ij}^A\tilde e_{ij}^A\rangle &=&
\left\{
\begin{array}{cc}
4\frac{\lambda_1\sqrt{\lambda_1^2+\lambda_2^2+\mu}-\sqrt{2}(\lambda_2^2+\mu)}
{\sqrt{\lambda_1^2\lambda_2^2+\mu(2\lambda_2^2+\lambda_1^2+2\mu)}} & \text{in phase 1} \\
4\frac{\lambda_1\sqrt{\lambda_1^2+\lambda_2^2+\mu}+\sqrt{2}(\lambda_2^2+\mu)}
{\sqrt{\lambda_1^2\lambda_2^2+\mu(2\lambda_2^2+\lambda_1^2+2\mu)}} & \text{in phase 2}\,,
\end{array}\right. \\
\n\\
\la{ff-1}
\langle f_{ij}^A\tilde f_{ij}^A\rangle &=&
\left\{
\begin{array}{cc}
4\frac{\sqrt{2}\lambda_2(\lambda_1+\sqrt{2}\sqrt{\lambda_1^2+\lambda_2^2+\mu})}
{\sqrt{\lambda_1^2\lambda_2^2+\mu(2\lambda_2^2+\lambda_1^2+2\mu)}} & \text{in phase 1} \\
4\frac{\sqrt{2}\lambda_2(-\lambda_1+\sqrt{2}\sqrt{\lambda_1^2+\lambda_2^2+\mu})}
{\sqrt{\lambda_1^2\lambda_2^2+\mu(2\lambda_2^2+\lambda_1^2+2\mu)}} & \text{in phase 2}\,,
\end{array}\right. \\
\n\\
\la{O-1}
\langle O\rangle \bydef \langle\(\psi^\dagger\psi\)^2\rangle &=&
\left\{
\begin{array}{cc}
\frac{-1}{2\sqrt{2}\sqrt{\lambda_1^2+\lambda_2^2+\mu}} & \text{in phase 1} \\
\frac{1}{2\sqrt{2}\sqrt{\lambda_1^2+\lambda_2^2+\mu}} & \text{in phase 2}\,.
\end{array}\right.
\eea

We now turn to constructing the mean field approximation to the model, to check its accuracy
against the exact calculation. We apply the general method of Section V, which is rather straightforward
in this simple case. We first take the ``cavity'' in the form of two neighbor lattice sites connected
by a link (1$^{\rm st}$ approximation), and then three adjacent sites connected by two links
(2$^{\rm nd}$ approximation). Both mean-field approximations give satisfactory accuracy when compared
to the exact results but the second is, of course, better.

In both cases the cavity boundary is just the neighbor sites connected to the cavity by
link variables $U_{mi}$. Expanding $e^{S_{mi}}$ up to the second power (higher powers are zero
because of too many fermion operators) and integrating over $U_{mi}$ we obtain several operator
structures. Splitting them into the product of operators composed of fields inside the chosen cavity,
and the operator built of the outside fields, we replace the latter by the averages, to be found
self-consistently. Most of the operators break the $\sigma^3$-symmetry of the original action, and
we ignore them. The only operator with proper symmetries left is $O=(\psi^\dagger_i\psi_i)^2$.
We find the effective action for the two-site cavity
\begin{eqnarray}
e^{S(m)}=1+\lambda_1\(\langle O\rangle +(\psi^\dagger_m\psi_m)^2\)
+2(\lambda^2_1+\lambda^2_2+\mu)\langle O\rangle(\psi^\dagger_m\psi_m)^2.
\la{Seff-D1}\end{eqnarray}
To obtain the self-consistency equation we equate the average of $O$ found from the effective action
\ur{Seff-D1}, to $\langle O\rangle$. The resulting nonlinear equation on $\langle O\rangle$ has
three solutions. We choose the solutions $\langle O\rangle(\lambda_1,\lambda_2,\mu)$ that are
real in the ranges 1 and 2 above.

The results for the averages of three operators in the first and second approximations as well
as their exact values \urss{ee-1}{ff-1}{O-1} are presented in Fig.~9. There is a ``phase transition''
between phases 1 and 2 at $\lambda_1=0$. We see that the second mean-field approximation corresponding
to a three-site, two-segment cavity gives a very satisfactory accuracy.

\begin{figure}[t]
\includegraphics[width=0.3\textwidth]{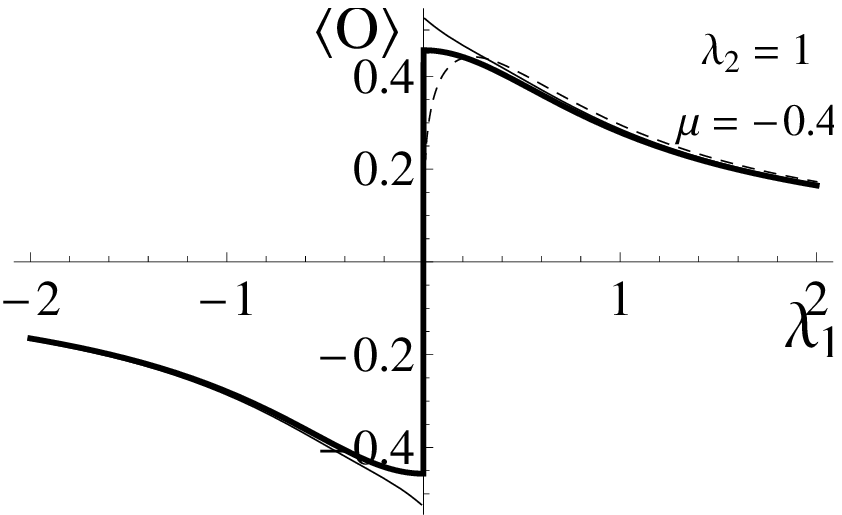}
\includegraphics[width=0.3\textwidth]{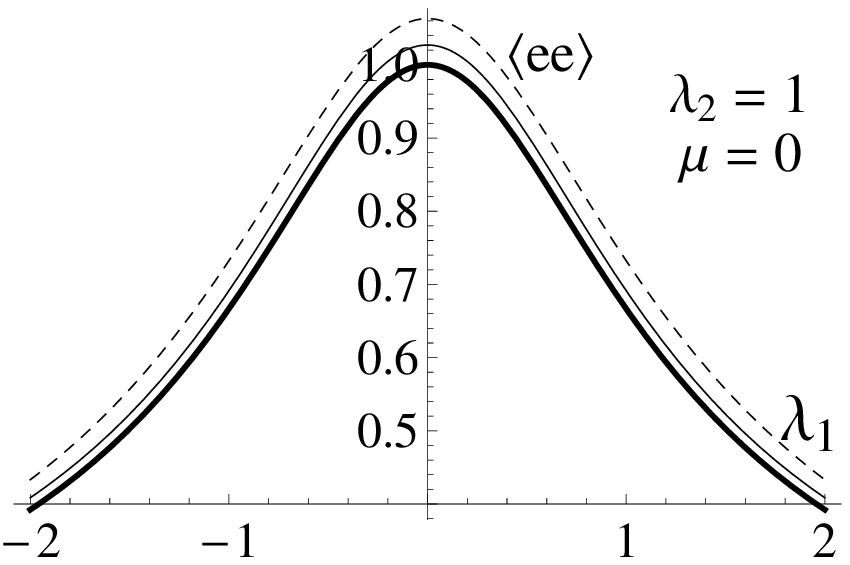}
\includegraphics[width=0.3\textwidth]{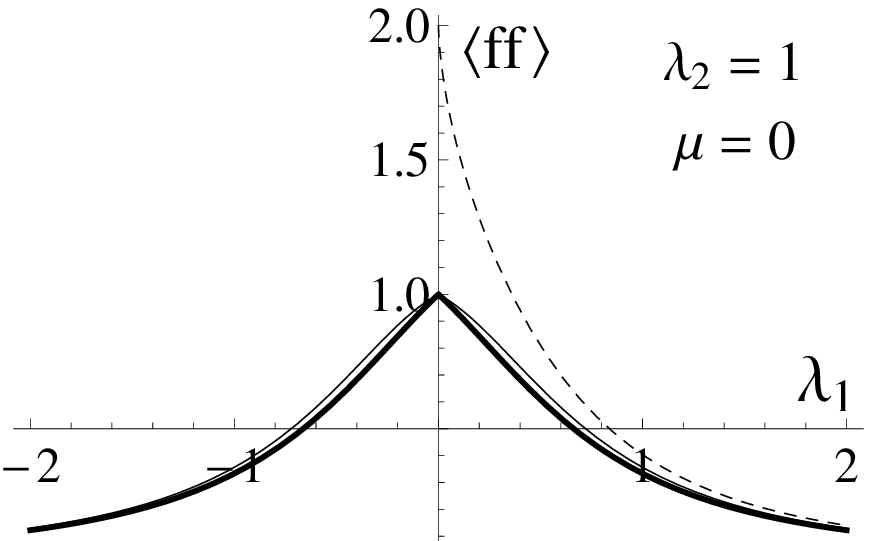}
\caption{Averages of three operators in the $1d$ model. The bold line is the exact result,
the dashed line is the result of the 1$^{\rm st}$ mean-field approximation using one lattice segment,
the solid line is the result of the 2$^{\rm nd}$ (two-segment) mean-field approximation.}
\end{figure}


\end{document}